\documentclass[useAMS,usenatbib]{mn2e}                                  
\usepackage{ amsmath }
\usepackage{ amssymb }
\usepackage{graphicx}
\usepackage{aas_macros}
\usepackage{deluxetable}
\usepackage{subfigure}
\begin{document}
\title[Mid-J CO Observations of Perseus B1-E5]{Mid-J CO observations of Perseus B1-East 5: evidence for turbulent dissipation via low-velocity shocks}

\author[A. Pon et al.] {Andy Pon, $^{1,2}$ Doug Johnstone, $^{3,4,5}$ Michael J. Kaufman, $^{6,7}$, Paola Caselli $^{1,2}$ 
\newauthor
and Ren\'{e} Plume$^{8}$\\
$^1$Max Planck Institute for Extraterrestrial Physics, Giessenbachstrasse 1, D-85748 Garching, Germany\\
$^2$School of Physics and Astronomy, University of Leeds, Leeds LS2 9JT, UK\\
$^3$Joint Astronomy Centre, 660 North A'ohoku Place, University Park, Hilo, HI 96720, USA\\
$^4$NRC-Herzberg Institute of Astrophysics, 5071 West Saanich Road, Victoria, BC V9E 2E7, Canada\\
$^5$Department of Physics and Astronomy, University of Victoria, PO Box 3055 STN CSC, Victoria, BC V8W 3P6, Canada\\
$^6$Department of Physics and Astronomy, San Jose State University, One Washington Square, San Jose, CA 95192-0106, USA\\
$^7$Space Science and Astrobiology Division, MS 245-3, NASA Ames Research Center, Moffett Field, CA 94035, USA\\
$^8$Department of Physics \& Astronomy, University of Calgary, Calgary, AB T2N 1N4, Canada}

\maketitle

\begin{abstract}
Giant molecular clouds contain supersonic turbulence and magnetohydrodynamic simulations predict that this turbulence should decay rapidly. Such turbulent dissipation has the potential to create a warm (T $\sim100$ K) gas component within a molecular cloud. We present observations of the CO J = 5-4 and 6-5 transitions, taken with the {\it Herschel Space Observatory}, towards the Perseus B1-East 5 region. We combine these new observations with archival measurements of lower rotational transitions and fit photodissociation region models to the data. We show that Perseus B1-E5 has an anomalously large CO J = 6-5 integrated intensity, consistent with a warm gas component existing within the region. This excess emission is consistent with predictions for shock heating due to the dissipation of turbulence in low-velocity shocks with the shocks having a volume filling factor of 0.15 per cent. We find that B1-E has a turbulent energy dissipation rate of $3.5 \times 10^{32}$ erg s$^{-1}$ and a dissipation time-scale that is only a factor of 3 smaller than the flow crossing time-scale.
\end{abstract}


\section{INTRODUCTION}
\label{introduction}

Molecular line observations of giant molecular clouds (GMCs) reveal linewidths much larger than expected from thermal broadening alone and, as such, it is believed that molecular clouds contain significant supersonic turbulent motions (e.g. \citealt{Larson81,Solomon87}). Magnetohydrodynamic simulations of molecular clouds containing supersonic turbulence predict that this turbulence should decay rapidly due to the presence of shocks, with the turbulence decaying of the order of a free-fall time at the driving scale \citep{Gammie96, MacLow98, Stone98, MacLow99, Padoan99, Ostriker01}. 

\citet{Pon12Kaufman} computed the structure of low-velocity shocks, 2 and 3 km s$^{-1}$, propagating into gas with densities around 1000 cm$^{-3}$, as would be expected for shocks generated by the typical turbulent motions within molecular clouds. They show that such low-velocity shocks dissipate the majority of their energy in CO rotational transitions, with the remainder of the energy going into magnetic field compression or H$_2$ pure rotational line emission. This H$_2$ emission, however, is only significant for the stronger shocks in the parameter range of \citet{Pon12Kaufman}. By scaling the shock models with the rate of turbulent energy dissipation expected from numerical simulations (see also \citealt{Basu01}) and then comparing these scaled shock models with photodissociation region (PDR) models from \citet{Kaufman99}, \citet{Pon12Kaufman} predict that mid-J CO rotational emission ($J_{upper} \ge 6$) from molecular clouds should come primarily from shocked gas. \citet{Pon12Kaufman} predict that the J = 5 $\rightarrow$ 4 transition should have equal contributions from shocked and unshocked gas while lower lines should be dominated by emission from unshocked gas. In other words, \citet{Pon12Kaufman} predict that mid-J CO lines should be brighter than expected from FUV illuminated gas alone and should have line ratios indicative of a higher temperature component. This shock emission would be radiated from myriad shocks existing throughout a molecular cloud, with each individual shock being very thin, relative to the cloud, and the total volume filling factor of the shocked gas only being a few tenths of a per cent \citep{Pon12Kaufman}.

These predictions, however, are only valid for regions of molecular clouds where the only significant heating sources are shocks, cosmic rays, and an interstellar radiation field (ISRF) around 1 Habing. In much higher radiation fields, or near energetic outflows with velocities of the order of tens of kilometres per second, the extra heating from these sources is expected to wash out any signature of heating from the decay of turbulence in low-velocity shocks. In very active star-forming regions, such as in starburst galaxies, the rate of mechanical energy dissipation can be large enough to globally heat a molecular cloud and excite high-J CO transitions, rather than only heating small volume fractions of the gas \citep{Kazandjian12,Kazandjian14}.

The Perseus molecular cloud is a nearby ($\sim$300 pc), low-mass star-forming region (see \citealt{Bally08Walawender} and references therein). Perseus B1-East (B1-E) is a 0.1 deg$^2$ clump within the Perseus molecular cloud with $\sim$100 M$_\odot$ of material \citep{Bachiller86, Sadavoy12}. While B1-E has a peak visual extinction greater than 5, indicating a peak column density above $10^{22}$ cm$^{-2}$, previous continuum observations show no sign of embedded young stellar objects (e.g., \citealt{Enoch06, Kirk06, Jorgensen07, Evans09, Sadavoy14}). The large column densities, coupled with the lack of embedded protostellar sources, make B1-E a prime example to test the predictions of \citet{Pon12Kaufman}, as there should be numerous shocks along any given line of sight but no obvious additional heating sources that could mimic the effects of shock heating, nor significant stellar luminosity to heat the clump gas.

\citet{Sadavoy12} observed continuum emission from B1-E with the {\it Herschel Space Observatory} ({\it Herschel}) and find that B1-E contains nine separate substructures existing in a lower density, extended environment. Each substructure has a mass of the order of a Solar mass, an average density of a few times 10$^4$ cm$^{-3}$, and a radius around 10 000 au. Based upon a 25 K upper limit on the kinetic temperature of these substructures, \citet{Sadavoy12} find that all but one of these structures are unbound (B1-E2), despite B1-E as a whole being gravitationally bound. 

The source denoted as B1-E5 by \citet{Sadavoy12} is centred at RA $3^h36^m37^s.3$ and Dec. $31^\circ11\arcmin41\arcsec$ (J2000). B1-E5 has a peak H$_2$ column density of $1.45 \times 10^{22}$ cm$^{-2}$ and is the most spatially extended of the nine substructures observed by \citet{Sadavoy12}, with an angular diameter of the order of an arcminute. B1-E5 has a radius of $9.3 \times 10^3$ au (0.05 pc), a mass between 0.5 and 2 Solar masses, an average density between 2.4 and 9.1 $\times 10^4$ cm$^{-3}$, and a gas temperature less than 20 K \citep{Sadavoy12}. NH$_3$ observations of B1-E5 yield a v$_{LSR} = 7.57$ km s$^{-1}$ and a full width at half-maximum (FWHM) of 0.833 km s$^{-1}$ \citep{Sadavoy12}.

Towards B1-E, there are three distinct velocity components at 3, 7.5, and 10 km s$^{-1}$ (e.g., \citealt{Sun06}; Sadavoy et al., in preparation). The 10 km s$^{-1}$ component only appears in the far north-east corner of B1-E and is not expected to be present in observations towards B1-E5. The 3 km s$^{-1}$ component can be seen throughout most of B1-E and on larger scales is only detected in the general vicinity of B1-E and in the very far southwestern portion of the Perseus molecular cloud. Finally, the 7.5 km s$^{-1}$ component is the strongest of the three components towards B1-E and is detected throughout the majority of the Perseus molecular cloud. See fig. 3 of \citet{Sun06} for velocity channel maps illustrating the spatial distribution of these components. The detection of multiple velocity components is common in both low-mass and high-mass star-forming regions (e.g., \citealt{Hacar13, Henshaw13}). For this paper, north refers to larger declination and east to larger right ascension.

In Section \ref{observations} of this paper, we present $^{12}$CO J = 6 $\rightarrow$ 5 and 5 $\rightarrow$ 4 observations centred on the B1-E5 substructure, taken with the Heterodyne Instrument for the Far-Infrared (HIFI) \citep{deGraauw10} on board the {\it Herschel Space Observatory} \citep{Pilbratt10}. We also present fits to lower J, archival CO data towards B1-E5 in Section \ref{archival}. In Section \ref{sed}, we attempt to fit spectral line energy distributions (SLEDs) to the CO observations, based upon PDR models from \citet{Kaufman99}, \textsc{meudon} PDR models \citep{LePetit06}, and \textsc{kosma}-$\tau$ PDR models \citep{Rollig06}. We discuss the quality of these SLED fits and suggest reasons for why there are discrepancies between the data and the models in Section \ref{discussion}. With the interpretation that excess emission is due to low-velocity shocks, we derive the properties of these shocks and the rate at which turbulence is being dissipated in Section \ref{shock}. Finally, we summarize our key findings in Section \ref{conclusions}.

\section{OBSERVATIONS}
\label{observations}

Observations of the $^{12}$CO J = 6 $\rightarrow$ 5 and 5 $\rightarrow$ 4 transitions were obtained using HIFI on the {\it Herschel Space Observatory} as part of the guaranteed time project GT2\_apon\_1. Single pointings were obtained at RA $3^h36^m37^s.76$ and Dec. $31^{\circ}11\arcmin39\arcsec.5$ (J2000), which corresponds to the centre of the substructure denoted as B1-E5 by \citet{Sadavoy12}. The {\it Herschel} beam at these wavelengths has a half power beamwidth of approximately 35 arcsec, roughly half the diameter of B1-E5. Frequency switch mode was used for both observations. 

HIFI has two back-end spectrometers, the Wide Band Spectrometer (WBS) and the High Resolution Spectrometer (HRS). Each spectrometer, in turn, measures the intensity of two perpendicular polarizations, denoted as H and V, for horizontal and vertical. The WBS has a fixed resolution of 1.1 MHz while the HRS has a variable resolution. A resolution setting of 1.0 MHz was used for the HRS for the 6 $\rightarrow$ 5 observations and a resolution of 0.5 MHz for the 5 $\rightarrow$ 4 observations. 

The CO J = 5 $\rightarrow$ 4 transition has a rest frequency of 576.2679305 GHz \citep{Remijan07}, an upper energy level of 83 K, and a critical density of $\sim 2 \times 10^5$ cm$^{-3}$ \citep{Schoier05}. The CO J = 5 $\rightarrow$ 4 observations were taken on 2012 July 31 as observation ID (OBSID) 1342248900 in HIFI band 1b. The on-source integration time was 321 s and the beam was 37 arcsec. A small positive frequency throw of 92.25 MHz was used. The spectral resolution was 0.26 km s$^{-1}$ for the WBS and 0.12 km s$^{-1}$ for the HRS. 

The CO J = 6 $\rightarrow$ 5 transition has a rest frequency of 691.473076 GHz \citep{Remijan07}, an upper energy level of 116 K, and a critical density of $3 \times 10^5$ cm$^{-3}$ \citep{Schoier05}. The CO J = 6 $\rightarrow$ 5 observations were taken on 2012 August 10 as observation ID (OBSID) 1342249401 in HIFI band 2a. The on-source integration time was 3845 s and the beam was 33 arcsec. A small positive frequency throw of 96.00 MHz was used. The spectral resolution was 0.22 km s$^{-1}$  for the WBS and 0.21 km s$^{-1}$ for the HRS. The main beam efficiency of HIFI is 0.75 for both observed transitions and all intensity values in this paper are given in main beam temperature (T$_{MB}$) units, unless otherwise stated. Table \ref{table:obssetup} contains a summary of the setup for each of the observations 

All data reduction was done using the {\it Herschel} Interactive Processing Environment (\textsc{hipe}) version 11.0.0 \citep{Ott10}. The data were first repipelined using the standard \textsc{hipe} pipeline included with version 11.0.0. Standing waves were removed from the data using the fitHifiFringe task, the remaining baselines were fit using the fitBaseline task, and the doFold task was used to fold the data to account for the frequency switching. Overlapping subbands in the HRS data were averaged together using the doStitch command. The V and H polarizations were then combined and regions at the edges of the spectra where the noise levels were elevated were removed. 

The CO J = 6 $\rightarrow$ 5 line and its negative counterpart appear in different, non-overlapping subbands in the HRS data and the \textsc{hipe} doFold task was unable to properly fold the data across two subbands. As such, the CO J = 6 $\rightarrow$ 5 HRS data are not dealt with further in this paper. This is of little concern since the WBS data have a better signal-to-noise ratio than the HRS data. 

In both the CO J = 6 $\rightarrow$ 5 and 5 $\rightarrow$ 4 spectra, two separate velocity components are obvious and will be hereafter referred to as the 3 and 8 km s$^{-1}$ components, based upon their approximate central, local standard of rest velocities. A 10 km s$^{-1}$ component is not obviously detected in any spectrum. Gaussian profiles were fit to both of these components and the uncertainty in the parameters of these fits were estimated using the \textsc{hipe} SpectrumFitter routine. The uncertainty in the integrated intensity was calculated via
\begin{equation}
\mbox{d}I = \mbox{rms} \times \mbox{velocity resolution} \sqrt{\frac{2 \times \mbox{FWHM}}{\mbox{velocity resolution}}},
\label{eqn:di}
\end{equation} 
where d$I$ is the uncertainty in the integrated intensity, rms is the root mean square of the baseline, and FWHM is the full width at half-maximum of the line. 

We were unable to fully remove the baseline fluctuations from the CO 6 $\rightarrow$ 5 data set, but have not included the strength of these fluctuations in the uncertainty estimate for the 6 $\rightarrow$ 5 line strength. The absolute flux calibration for these HIFI bands is of the order of 10 per cent \citep{Roelfsema12}, but this calibration uncertainty has not been included in any stated intensity uncertainties. 

The frequency accuracies of both spectrometers are well below the frequency resolution \citep{Roelfsema12}, such that this intrinsic frequency calibration uncertainty is negligible. The pointing accuracy of {\it Herschel} is of the order of an arcsecond or less, and thus is negligible compared to the beam size \citep{Roelfsema12}. 

In the 5 $\rightarrow$ 4 data, the 8 km s$^{-1}$ component shows some substructure in both the WBS and HRS data, although this deviation from a single Gaussian profile is more obvious in the HRS data. The FWHM, peak intensity, and central line velocity of the 8 km s$^{-1}$ component were obtained by fitting the 8 km s$^{-1}$ component with only one Gaussian, but the total integrated intensity was obtained by fitting two Gaussians to this component and then summing the integrated intensities of the two Gaussians. The integrated intensity of the 8 km s$^{-1}$ component as calculated from the single Gaussian fit, however, only differs by 2 per cent from the value obtained by fitting two Gaussians to the 8 km s$^{-1}$ component. Lower J $^{12}$CO and $^{13}$CO lines (see Section \ref{archival}) do not show any obvious signs of this substructure in the 8 km s$^{-1}$ component, although most, if not all, of these lower J lines are optically thick and have flat tops. We thus treat the 8 km s$^{-1}$ component as a single component for the remainder of this paper.

\begin{figure}
   \centering
   \includegraphics[width=3in]{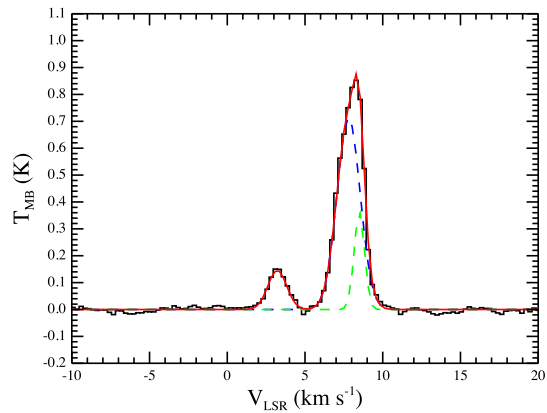}
   \caption{CO J = 5 $\rightarrow$ 4 spectrum of B1-E5, as detected by the WBS, is shown as the black histogram. The overall best fit to the data is shown as the solid red line, and the two different Gaussian fits used to fit the 8 km s$^{-1}$ component are shown as the blue and light green dashed lines.}
   \label{fig:co54wbs}
\end{figure}

\begin{figure}
   \centering
   \includegraphics[width=3in]{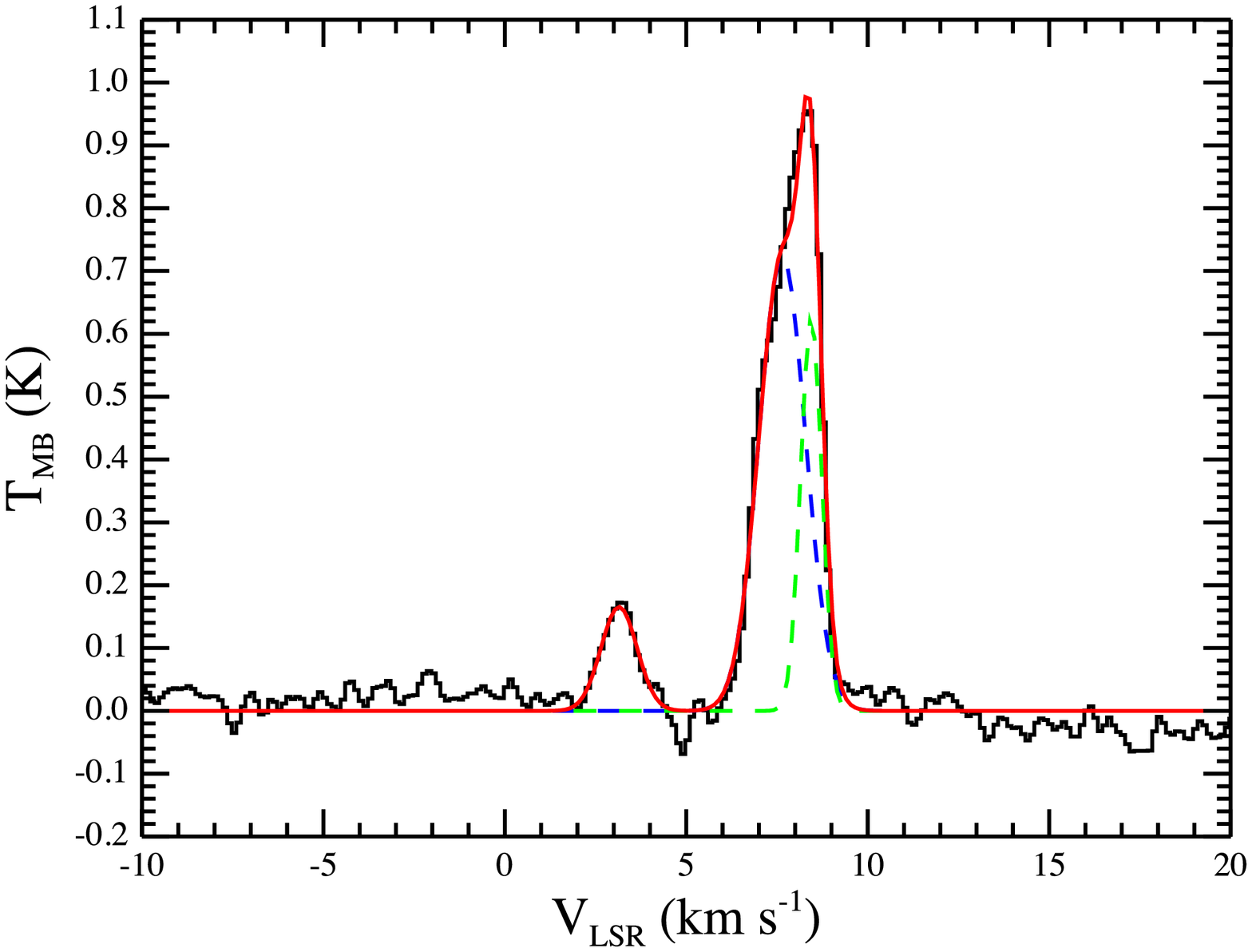}
   \caption{CO J = 5 $\rightarrow$ 4 spectrum of B1-E5, as detected by the HRS, is shown as the black histogram. The overall best fit to the data is shown as the solid red line, and the two different Gaussian fits used to fit the 8 km s$^{-1}$ component are shown as the blue and light green dashed lines.}
   \label{fig:co54hrs}
\end{figure}

\begin{figure}
   \centering
   \includegraphics[width=3in]{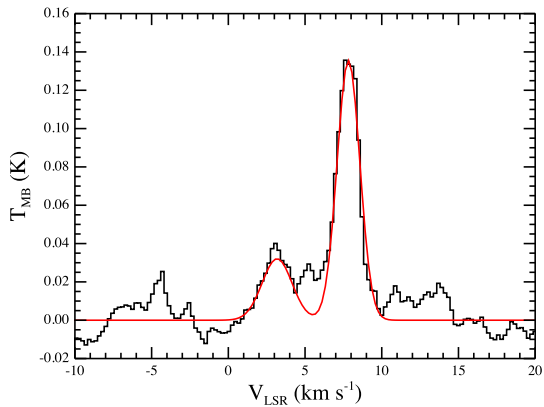}
   \caption{CO J = 6 $\rightarrow$ 5 spectrum of B1-E5, as detected by the WBS, is shown as the black histogram. The overall best fit to the data is shown as the solid red line.}
   \label{fig:co65wbs}
\end{figure}

\begin{figure}
   \centering
   \includegraphics[width=3in]{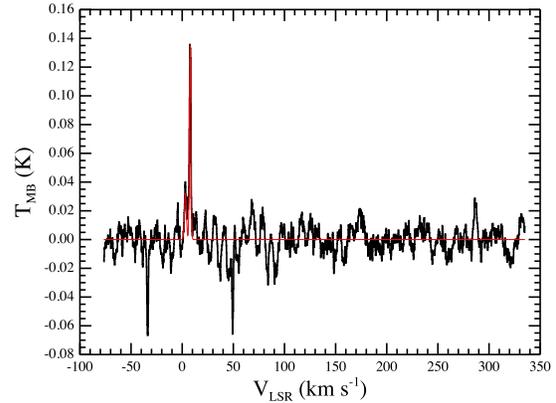}
   \caption{Full velocity range of the CO J = 6 $\rightarrow$ 5 spectrum of B1-E5, as detected by the WBS, is shown as the black histogram. The overall best fit to the data is shown as the solid red line. The negative features at velocities near -35 and 50 km s$^{-1}$ are ghost images of the main line created from the frequency switching mode that the observations were taken with.}
   \label{fig:co65wbsfull}
\end{figure}

The best-fitting profiles are shown with the reduced spectra in Figs \ref{fig:co54wbs}-\ref{fig:co65wbs}. Figs \ref{fig:co54wbs} and \ref{fig:co54hrs} also show the two Gaussian components fit to the 8 km s$^{-1}$ component. The full spectral range observed around the CO J = 6 $\rightarrow$ 5 line is shown in Fig. \ref{fig:co65wbsfull}, to better illustrate the residual baseline noise that we were unable to fully remove. Table \ref{table:fits} contains the details of the best fits. 

\begin{table*}
\begin{minipage}{\textwidth}
\caption{Observational setup.}
\begin{center}
\begin{tabular}{cccccccccc}
\hline
\hline
Line & Spectrometer & Frequency & Band & T$_{\mbox{int}}$ & Resolution & HPBW & $E_{\mbox{up}}$ & $n_{\mbox{crit}}$ & $\eta_{\mbox{MB}}$ \\
& & (GHz) & & (s) & (km s$^{-1}$) & (arcsec) & (K) & (10$^5$ cm$^{-3}$) & \\
(1) & (2) & (3) & (4) & (5) & (6) & (7) & (8) & (9) & (10) \\
\hline
$^{12}$CO J = 5 $\rightarrow$ 4 & WBS & 576.2679305 & 1b & 328 & 0.26 & 37 & 83 & 2 & 0.75\\
$^{12}$CO J = 5 $\rightarrow$ 4 & HRS & 576.2679305 & 1b & 328 & 0.12 & 37 & 83 & 2 & 0.75\\
$^{12}$CO J = 6 $\rightarrow$ 5 & WBS & 691.473076 & 2a & 3981 & 0.22 & 31 & 116 & 3 & 0.75\\
$^{12}$CO J = 6 $\rightarrow$ 5 & HRS & 691.473076 & 2a & 3981 & 0.21 & 31 & 116 & 3 & 0.75\\
\hline
\end{tabular}
\tablecomments{Column 1 gives the line observed and column 2 gives the spectrometer used. The rest frequency of the line is given in column 3 and the {\it Herschel} band that the line lies within is given in column 4. Column 5 gives the duration of the observation and column 6 gives the velocity resolution. Column 7 gives the half power beam width of the beam, column 8 gives the upper energy level of the transition, and column 9 gives the critical density of the transition. Columns 8 and 9 are based upon data in the Leiden Atomic and Molecular Database (LAMDA; \citealt{Schoier05}). The main beam efficiency is given in column 10. }
 \label{table:obssetup}
\end{center}
\end{minipage}
\end{table*}

\begin{table*}
\begin{minipage}{\textwidth}
\caption{Gaussian fitting results.}
\begin{center}
\begin{tabular}{cccccccccc}
\hline
\hline
Line & $V$ & d$V$ & $T_{MB}$ & d$T_{MB}$ & FWHM & dFWHM & $I$ & d$I$ & rms \\
 & (km s$^{-1}$) & (km s$^{-1}$) & (K) & (K) & (km s$^{-1}$) & (km s$^{-1}$) & (K km s$^{-1}$) & (K km s$^{-1}$) & (K) \\
(1) & (2) & (3) & (4) & (5) & (6) & (7) & (8) & (9) & (10)\\
\hline
WBS $^{12}$CO 5 $\rightarrow$ 4 & 3.22 & 0.10 & 0.15 & 0.02 & 1.42 & 0.23 & 0.220 & 0.008 & 0.010 \\
			            & 7.99 & 0.02 & 0.85 & 0.02 & 1.77 & 0.04 & 1.635 & 0.009 & 0.010 \\
HRS $^{12}$CO  5 $\rightarrow$ 4 & 3.15 & 0.12 & 0.17 & 0.03 & 1.17 & 0.27 & 0.207 & 0.015 & 0.027\\
				   & 7.95 & 0.03 & 0.91 & 0.03 & 1.68 & 0.06 & 1.615 & 0.017 & 0.027\\
WBS $^{12}$CO 6 $\rightarrow$ 5 & 3.13 & 0.15 & 0.032 & 0.005 & 2.29 & 0.36 & 0.078 & 0.009 & 0.009\\
				    & 7.83 & 0.03 & 0.136 & 0.005 & 1.75 & 0.08 & 0.253 & 0.008 & 0.009\\
\hline
\end{tabular}
\tablecomments{Column 1 gives the rotational transition of CO and the spectrometer data set that is fit. Columns 2 and 3 give the central velocity of the line, with respect to the local standard of rest, and the uncertainty in the central velocity. The peak intensity of the line and the uncertainty in this peak intensity are given in columns 4 and 5 in units of T$_{MB}$. The FWHM of the line is given in column 6 and the uncertainty in the FWHM is given in column 7. Columns 8 and 9 give the integrated intensity of the line and the associated uncertainty, while column 10 gives the root mean square of the continuum baseline away from the line.}
\label{table:fits}
\end{center}
\end{minipage}
\end{table*}

Comparing the WBS and HRS CO J = 5 $\rightarrow$ 4 fits, it is evident that the two overall fits are reasonably similar. The derived integrated intensities are the same to within errors, while the HRS data have a slightly higher peak intensity and slightly smaller FWHM. The noise level of the HRS data, however, is more than twice the noise level of the WBS data. The WBS and HRS data also share many of the same noise sources, such that the noise between the two spectra is correlated. Thus, we do not combine the HRS data and the WBS data and elect to only use the results of the fits to the WBS data in the remainder of this paper.

\section{Archival Data}
\label{archival}
While the CO J = 6 $\rightarrow$ 5  and 5 $\rightarrow$ 4 lines are predicted to have a significant component from shocks \citep{Pon12Kaufman}, observations of lower CO lines are needed to constrain the properties of the ambient, unshocked gas towards B1-E5, in order to estimate the amount of emission in the higher CO lines coming from this ambient gas. 

\subsection{The COMPLETE survey}
\label{complete}

The Coordinated Molecular Probe Line Extinction Thermal Emission Survey of Star Forming Regions (COMPLETE survey; \citealt{Ridge06DiFrancesco}) used the Five Colleges Radio Astronomy Observatory (FCRAO) 14 metre telescope to survey the entire Perseus molecular cloud in the $^{12}$CO J = 1 $\rightarrow$ 0 line at 115.27120 GHz and the $^{13}$CO J = 1 $\rightarrow$ 0 line at 110.20135 GHz. The beam for the COMPLETE observations was roughly 47 arcsec, slightly larger than but comparable to the {\it Herschel} beams. The COMPLETE beams are also still smaller than the diameter of B1-E5. Details regarding these observations are shown in Table \ref{table:otherlines}.

The COMPLETE data were downloaded and, for both isotopologues, the spectrum located closest to where the {\it Herschel} observations were conducted was extracted. For both isotopologues, the closest spectrum is centred at RA $3^h36^m37^s.7$ and Dec. $31^{\circ}11\arcmin47\arcsec$ (J2000), just 7 arcsec from the centre of the {\it Herschel} pointings. The data were converted from antenna temperature (T$_A^*$) units to main beam temperature units by applying an efficiency factor of 0.45 to the $^{12}$CO data and an efficiency factor of 0.49 to the $^{13}$CO data.

Both $^{12}$CO and $^{13}$CO spectra show components at roughly 3 and 8 km s$^{-1}$, as seen in the {\it Herschel} data. The lines have relatively flat tops, such that the integrated intensities of the lines were calculated by summing the emission over the channels associated with each component, rather than by trying to fit Gaussian profiles. The error in the integrated intensity was calculated from Equation \ref{eqn:di}, where the full range over which the emission was summed was used instead of twice the FWHM. A central velocity was estimated from the flux weighted mean of each component and the uncertainty in the central velocity was derived by taking the difference between the central velocity from a Gaussian fit to the line in \textsc{idl} and the flux weighted mean velocity. The maximum intensity was picked as the peak amplitude and the error in the maximum intensity was set to be equal to the rms of the baseline. The FWHM was found from the width at the intensity of half of the peak and, since the lines have sharp sides, the uncertainty in the FWHM was estimated as twice the velocity resolution, since each edge of the profile may be off by one velocity element. The full parameter set derived from this data is shown in Table \ref{table:otherfits}.
 
\begin{table*}
\begin{minipage}{\textwidth}
\caption{Archival line data sources.}
\begin{center}
\begin{tabular}{ccccccccccc}
\hline
\hline
Line & Telescope & RA & Declination & Offset & Frequency & Resolution & HPBW & $E_{\mbox{up}}$ & $n_{\mbox{crit}}$ & $\eta_{\mbox{MB}}$ \\
 & & (h:m:s) & ($^{\circ}:\arcmin:\arcsec$) & (arcsec) & (GHz) & (km s$^{-1}$) & (arcsec) & (K) & (10$^5$ cm$^{-3}$) & \\
 (1) & (2) & (3) & (4) & (5) & (6) & (7) & (8) & (9) & (10) & (11) \\
\hline
$^{12}$CO J = 1 $\rightarrow$ 0 & FCRAO & 3:36:37.7 & 31:11:47 & 7 & 115.27120 & 0.064 & 47 & 6 & 0.02 & 0.45\\
$^{13}$CO J = 1 $\rightarrow$ 0 & FCRAO & 3:36:37.7 & 31:11:47 & 7 &110.20135 & 0.066 & 49 & 5 & 0.02 & 0.49\\
$^{13}$CO J = 2 $\rightarrow$ 1 & KOSMA & 3:36:37.4 & 31:11:52 & 14 & 220.39868 & 0.225 & 114 & 16 & 0.1 & 0.68\\
$^{12}$CO J = 3 $\rightarrow$ 2 & KOSMA & 3:36:37.4 & 31:11:52 & 14 & 345.79599 & 0.295 & 73 & 33 & 0.4 & 0.70\\
\hline
\end{tabular}
\tablecomments{Column 1 gives the line observed and column 2 gives the telescope used. The right ascension and declination of the observation are given in columns 3 and 4, respectively, while the angular offset from the central position of the {\it Herschel} observations is given in column 5. The rest frequency of the line is given in column 6 and column 7 gives the velocity resolution. Column 8 gives the half power beamwidth of the beam, column 9 gives the upper energy level of the transition, and column 10 gives the critical density of the transition. Columns 9 and 10 are based upon data in the Leiden Atomic and Molecular Database (LAMDA; \citealt{Schoier05}). The main beam efficiency is given in column 11.}
 \label{table:otherlines}
\end{center}
\end{minipage}
\end{table*}

\begin{table*}
\begin{minipage}{\textwidth}
\caption{Archival line properties.}
\begin{center}
\begin{tabular}{cccccccccc}
\hline
\hline
Line & $V$ & d$V$ & $T_{MB}$ & d$T_{MB}$ & FWHM & dFWHM & $I$ & d$I$ & rms \\
 & (km s$^{-1}$) & (km s$^{-1}$) & (K) & (K) & (km s$^{-1}$) & (km s$^{-1}$) & (K km s$^{-1}$) & (K km s$^{-1}$) & (K) \\
(1) & (2) & (3) & (4) & (5) & (6) & (7) & (8) & (9) & (10)\\
\hline
$^{12}$CO 1 $\rightarrow$ 0 & 3.49 & 0.15 & 8.18 & 0.18 & 2.35 & 0.13 & 18.7 & 0.2 & 0.18 \\
			                     & 7.78 & 0.05 & 12.89 & 0.18 & 2.29 & 0.13 & 31.8 & 0.2 & 0.18 \\
$^{13}$CO 1 $\rightarrow$ 0 & 3.36 & 0.02 & 3.00 & 0.04 & 1.73 & 0.13 & 3.94 & 0.10 & 0.04 \\
			                     & 7.73 & 0.07 & 6.65 & 0.04 & 1.86 & 0.13 & 12.90 & 0.12 & 0.04 \\
$^{13}$CO 2 $\rightarrow$ 1 & 3.07 & 0.09 & 1.25 & 0.36 & 0.92 & 0.23 & 1.2 & 0.3 & 0.36\\
			                     & 7.69 & 0.08 & 2.69 & 0.36 & 2.68 & 0.20 & 7.7 & 0.4 & 0.36\\
$^{12}$CO 3 $\rightarrow$ 2 & 3.13 & 0.08 & 3.71 & 0.65 & 1.6 & 0.2 & 6.4 & 0.7 & 0.65\\
			                     & 7.73 & 0.06 & 6.85 & 0.65  & 2.6 & 0.1 & 19.0 & 0.8 & 0.65\\
\hline
\end{tabular}
\tablecomments{Column 1 gives the line measured. Columns 2 and 3 give the central velocity of the line, with respect to the local standard of rest, and the uncertainty in the central velocity. The peak intensity of the line and the uncertainty in this peak intensity are given in columns 4 and 5 in main beam temperature units. The FWHM of the line is given in column 6 and the uncertainty in the FWHM is given in column 7. Columns 8 and 9 give the integrated intensity of the line and the associated uncertainty, while column 10 gives the root mean square of the continuum baseline away from the line.}
\label{table:otherfits}
\end{center}
\end{minipage}
\end{table*}

\subsection{KOSMA Data}
\label{kosma}

\citet{Sun06} used the K\"{o}lner Observatorium f\"{u}r SubMillimeter Astronomie (KOSMA) 3 metre telescope to survey the Perseus molecular cloud in the $^{12}$CO J = 3 $\rightarrow$ 2 and $^{13}$CO J =  2 $\rightarrow$ 1 transitions. The KOSMA beam for the $^{12}$CO observations was approximately 73 arcsec and the beam was 114 arcsec for the $^{13}$CO observations. While the $^{12}$CO observations have a beam approximately the same size as B1-E5, the $^{13}$CO observations have a beam larger than B1-E5. Therefore, there might be some beam dilution in the $^{13}$CO 2 $\rightarrow$ 1 data, such that the integrated intensities of the $^{13}$CO J = 2 $\rightarrow$ 1 line should be considered lower limits. 

The nearest KOSMA pointing to the {\it Herschel} pointings is one at R.A. $3^h36^m37^s.42$ and Dec. $31^\circ11\arcmin52\arcsec.29$ (J2000), an angular distance of 14 arcsec from the {\it Herschel} pointings. The details of the observations are given in Table \ref{table:otherlines}.

The 3 and 8 km s$^{-1}$ components are both visible in the KOSMA data and the \textsc{continuum and line analysis single-dish software}'s (\textsc{class}'s) Gaussian fitting routine was used to fit Gaussians to the lines, as well as to estimate the errors in all of the derived parameters. The uncertainty in the peak intensity was set to be the rms of the baseline, as before. The details of the fits are given in Table \ref{table:otherfits}.

\subsection{IRAM 30m Data}
\label{sadavoy}

Sadavoy et al. (in preparation) used the Institut de Radioastronomie Millim\'{e}trique (IRAM) 30 metre telescope to obtain maps of the $^{13}$CO J = 1 $\rightarrow$ 0, C$^{18}$O J = 1 $\rightarrow$ 0, and C$^{18}$O J = 2 $\rightarrow$ 1 lines around Perseus B1-E. The IRAM 30 m telescope has a resolution of $\sim12$ arcsec for the 2 $\rightarrow$ 1 line and 25 arcsec for the 1 $\rightarrow$ 0 lines. Sadavoy et al. (in preparation) also obtained maps of B1-E in the 2 $\rightarrow$ 1 transition of $^{12}$CO, $^{13}$CO, and C$^{18}$O, all at a resolution of $\sim35$ arcsec, using the Submillimeter Telescope (SMT). These data are not public yet, but an early analysis of the data indicates a $^{13}$CO J = 2 $\rightarrow$ 1 total integrated intensity of 8.6 K km s$^{-1}$ towards B1-E5 (Sadavoy, private communication), consistent with the combined 8.9 K km s$^{-1}$ integrated intensity from the KOSMA data, despite the much larger KOSMA beam. This suggests that the low J CO emission, at least, is fairly spatially uniform over the B1-E5 region and that the integrated intensities from the {\it Herschel} observations can be compared to the low J integrated intensities obtained with larger beam sizes.

\subsection{Comparison of Data Sets}
\label{comparison of data sets} 

The linewidth of the $^{12}$CO lines tends to decrease slightly with increasing J number, although all of the lines are within a factor of 2 in width. The low signal-to-noise ratio of the $^{12}$CO J = 6 $\rightarrow$ 5, 3 km s$^{-1}$ component is likely responsible for the slightly larger FWHM of this line. The central velocities of the $^{12}$CO lines vary by at most 0.36 km s$^{-1}$, suggesting that roughly the same gas is being traced by each component in the $^{12}$CO spectra. The $^{13}$CO central velocities and linewidths generally agree with the $^{12}$CO values, with the $^{13}$CO widths being slightly smaller than the $^{12}$CO widths, suggesting that the two different isotopologues are tracing approximately the same gas.

\begin{figure}
   \centering
   \includegraphics[width=3in]{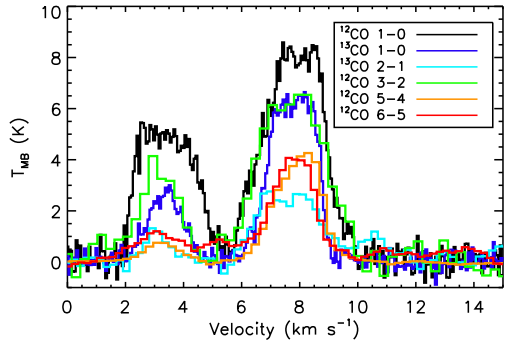}
   \caption{Spectra of various CO lines observed towards B1-E5. The black, green, orange, and red lines correspond to the J = 1 $\rightarrow$ 0, 3 $\rightarrow$ 2, 5 $\rightarrow$ 4, and 6 $\rightarrow$ 5 transitions of $^{12}$CO, while the dark blue and light blue lines correspond to the J = 1 $\rightarrow$ 0 and 2 $\rightarrow$ 1 transitions of $^{13}$CO. The J = 5 $\rightarrow$ 4 and 6 $\rightarrow$ 5 intensities have been scaled up by factors of 7 and 50, respectively, while the $^{12}$CO J = 1 $\rightarrow$ 0 intensities have been scaled down by a factor of 1.5.}
   \label{fig:overlay}
\end{figure}

Fig. \ref{fig:overlay} shows the two {\it Herschel} WBS spectra, the FCRAO spectra, and the KOSMA spectra all overlaid. The J = 5 $\rightarrow$ 4 and 6 $\rightarrow$ 5 intensities have been scaled up by factors of 5 and 30, respectively, while the $^{12}$CO J = 1 $\rightarrow$ 0 intensities have been scaled down by a factor of 1.5. There are no obvious differences in the line profiles, aside from the above mentioned differences in width. 

\section{CO SLED FITTING}
\label{sed}

While shocked gas will radiate strongly in CO lines, the unshocked, ambient gas in a molecular cloud will also contribute emission to CO rotational lines. To confirm the presence of shock heating, the emission from the ambient gas must first be removed from the data set. The only heating sources that should be significant for the ambient gas in Perseus B1-E5 are the incident ISRF and cosmic ray heating. Thus, PDR models, which include cosmic ray heating, should be able to explain the emission coming from the unshocked gas towards B1-E5. We compare our observations to the results of three different sets of PDR models: the \citet{Kaufman99} PDR models, \textsc{kosma}-$\tau$ models \citep{Sternberg89, Rollig06, Rollig13Szczerba}, and \textsc{meudon} models \citep{LePetit06}. Because different codes are known to give slightly different predictions (e.g., \citealt{Rollig07}), we are interested in any lines that are consistently under- or over predicted by the PDR models. 

\subsection{Kaufman et al. models}
\label{kaufman models}

The \citet{Kaufman99} PDR models are run for semi-infinite, constant density slabs of material illuminated on one side. The models use a microturbulent doppler linewidth of 1.5 km s $^{-1}$, which would correspond to an FWHM of 2.5 km s$^{-1}$ that is just slightly larger than the observed linewidths of B1-E5. In comparing the PDR model output with the observations of B1-E5, it is assumed that the PDR emission fills the beam. While the \citet{Kaufman99} models handle the chemistry of most common molecules quite well, they do not self-consistently calculate the self-shielding of $^{13}$CO. Therefore, only the $^{12}$CO data will be compared to the \citet{Kaufman99} PDR models. PDR models with an ISRF of log($G_0$) = -0.5, 0.5, and 1.5 ($G_0$ = 0.31, 3.1, and 31 Habing) are selected, where the average far-ultraviolet ISRF in free space is 1.7 Habing or $1.6 \times 10^{-3}$ erg cm$^{-2}$ s$^{-1}$ \citep{Tielens05}. For a low-mass star-forming region like B1-E, the local ISRF is not expected to be significantly above this average, free space value. Models with H nuclei densities of 10$^2$, 10$^3$, and 10$^4$ cm$^{-3}$ are selected. 

Observations suggest that CO is highly frozen on to dust grains at densities above 10$^5$ cm$^{-3}$ \citep{Caselli99}, while B1-E5 only has an average density of the order of 10$^4$ cm$^{-3}$ \citep{Sadavoy12}. Therefore, it is unlikely that CO is significantly frozen out of the gas phase in B1-E5. 

The combined $^{12}$CO J = 1 $\rightarrow$ 0 integrated intensity of the 3 and 8 km s$^{-1}$ components, from the FCRAO data, along with an $X_{\mbox{CO}}$ value of $2 \times 10^{20}$ cm$^{-2}$ K$^{-1}$ s (e.g., \citealt{Glover11}) suggest that the column density of H$_2$ towards B1-E5 should be of the order of $1 \times 10^{22}$ cm$^{-2}$. From dust continuum emission, \citet{Sadavoy12} estimate that the peak H$_2$ column density towards B1-E5 is $1.45 \times 10^{22}$ cm$^{-2}$. The depletion fraction, $f_D$, is typically defined as the ratio of the column density derived from continuum observations, assuming a typical gas-to-dust ratio, to the column density derived from CO observations and a reference CO fractional abundance. Depletion factors of 5-10 are commonly found in dense cores (e.g., \citealt{Caselli99, Hernandez11Caselli}), with depletion factors up to 78 being found towards infrared dark clouds \citep{Fontani12}. The $^{12}$CO J = 1 $\rightarrow$ 0 data indicate that the depletion factor for B1-E5 is approximately 1.5, suggesting that very little CO has frozen out of the gas phase in B1-E5. Since the \citet{Kaufman99} PDR code uses a steady state chemistry, PDR models that do not include freeze out are selected.

The $^{13}$CO observations can also be used to calculate the CO depletion factor. Under the assumption that the $^{12}$CO J = 1 $\rightarrow$ 0 line is highly optically thick, the excitation temperature of the gas is given by
\begin{equation}
T_{ex} = T_u \left[\mbox{ln}\left(\frac{T_u}{T_R + \frac{T_u}{\mbox{exp}\left(T_u/T_{bg}\right)-1}} + 1\right)\right]^{-1},
\end{equation}
where $T_{ex}$ is the excitation temperature of the emitting gas, $T_{R}$ is the observed peak intensity of the line, $T_{bg}$ is the 2.7 K background temperature, and $T_u$ is the upper energy level of the molecular line when expressed in units of kelvin. Since both the 3 and 8 km s$^{-1}$ $^{12}$CO J = 1 $\rightarrow$ 0 lines are flat topped, it is likely that both lines are optically thick, such that the excitation temperature for the two components would be 11 and 16 K, with the 3 km s$^{-1}$ component being the cooler component. 

Assuming that the $^{13}$CO lines are optically thin and that the excitation temperature for these lines is that derived from the $^{12}$CO J = 1 $\rightarrow$ 0 line, the column density of $^{13}$CO can be derived using
\begin{equation}
\begin{split}
N =& \frac{8\pi W \nu^2 k} {c^3 A h} \frac{g_l}{g_u} \left(\frac{1}{\mbox{exp}\left(\frac{h\nu}{kT_{ex}}\right)-1} - \frac{1}{\mbox{exp}\left(\frac{h\nu}{kT_{bg}}\right)-1}\right)^{-1} \\
&\times \frac{1}{\left(1 - \mbox{exp}\left(\frac{-h\nu}{kT_{ex}}\right)\right)} \frac{Q}{g_l \exp\left(\frac{-E_l}{kT_{ex}}\right)},
\end{split}
\label{eqn:column}
\end{equation}
where $W$ is the integrated intensity of the line, $A$ is the Einstein A coefficient for the transition, $h$ is Planck's constant, $g_l$ is the quantum weight of the lower transition, $g_u$ is the quantum weight of the upper transition where $g_J = 2J + 1$, $E_l$ is the lower energy level of the transition, and $Q$ is the partition function defined as $Q = \sum\limits_{j=0}^\infty (2J + 1) \exp\left(-E_J / kT\right)$. 

Given a number abundance of H$_2$ relative to $^{13}$CO of 4 $\times$ 10$^{5}$ \citep{Goodman09Pineda}, the $^{13}$CO J = 1 $\rightarrow$ 0 integrated intensity suggests that B1-E5, between both the 3 and 8 km s $^{-1}$ components, has a total column density of $1.1 \times 10^{22}$ cm$^{-2}$, consistent with the column density estimate from the $^{12}$CO J = 1 $\rightarrow$ 0 line. The $^{13}$CO 2 $\rightarrow$ 1 intensities, however, suggest a lower column density of $2 \times 10^{21}$ cm$^{-2}$, which would correspond to a depletion factor of the order of 10. This calculation, however, assumes that the level populations are in local thermodynamic equilibrium, and since the critical density of the $^{13}$CO J = 2 $\rightarrow$ 1 line is roughly 10$^{4}$ cm$^{-3}$, compared to 10$^{3}$ cm$^{-3}$ for the 1 $\rightarrow$ 0 line, it is likely that the lower depletion factor calculated for the 2 $\rightarrow$ 1 line is just due to non-LTE (local thermodynamic equilibrium) effects. A \textsc{radex} \citep{Vandertak07} model run with a background temperature of 2.73 K, a kinetic temperature of 15 K, an H$_2$ density of 10$^3$ cm$^{-3}$, a $^{13}$CO column density of $3.6 \times 10^{16}$ cm$^{-2}$, and an FWHM of 2.35 km s$^{-1}$ produces integrated intensities of the $^{13}$CO J = 1 $\rightarrow$ 0 and 2 $\rightarrow$ 1 lines that are both within 10 per cent of the observed integrated intensities, confirming that non-thermal excitation due to subcritical densities can explain the apparent weak emission in the $^{13}$CO J = 2 $\rightarrow$ 1 line.

The PDR models are for slabs of material illuminated on only one side, such that models with column densities roughly half of the molecular hydrogen column density of B1-E5 are selected and the intensity doubled for any line with an optical depth less than 0.5, to account for emission coming from the back half of the cloud. B1-E5 has a peak column density of $\sim1.45 \times 10^{22}$ cm$^{-2}$, which corresponds to an A$_V$ of 14 \citep{Sadavoy12}. By doubling the integrated intensities of the higher J CO lines, we make the highest possible estimate of the contribution from PDR gas; as a result, we are relatively conservative about the amount of emission that cannot be explained by PDRs.

CO integrated intensities are not strongly dependent upon the column density of the PDR model because most of the CO emission comes from the outer regions of the PDR. For higher J lines, larger gas temperatures are required to populate the higher rotational lines, such that little high-J emission comes from the cool gas in the interior of a cloud. Lower J lines of $^{12}$CO, on the other hand, tend to be optically thick, such that increasing the column density does not significantly affect the total integrated intensity of a low-J line. In all of the \citet{Kaufman99} PDR models examined, the $^{12}$CO J = 5 $\rightarrow$ 4 line is optically thick while the $^{12}$CO J = 6 $\rightarrow$ 5 line is optically thin ($\tau < 1$).

\subsection{MEUDON PDR models}
\label{meudon}

The \textsc{meudon} PDR code \citep{LePetit06} calculates the emission coming from a uniform density, plane parallel slab illuminated on two sides. The \textsc{meudon} PDR code computes the self-shielding of CO and its isotopologues using the approximation of \citet{Federman79}, rather than scaling the $^{13}$CO shielding from the $^{12}$CO shielding, as is done in the \citet{Kaufman99} code. Thus, the \textsc{meudon} code can be used to see if the observed low-J $^{13}$CO lines are consistent with a PDR model that fits the low-J $^{12}$CO lines. 

The \textsc{meudon} PDR code is publicly available and we run four models, all with a typical ISRF based upon \citet{Mathis83} and a total visual extinction (A$_V$) of 15 mag. The hydrogen nuclei density of the models is varied from 10$^2$ to 10$^5$ cm$^{-3}$ in even, logarithmic steps. All models are run with a turbulent doppler broadening velocity of 1 km s$^{-1}$, which corresponds to an FWHM of approximately 1.7 km s$^{-1}$, consistent with the observed FWHM of lines from B1-E5. 

\subsection{KOSMA-$\tau$}
\label{kosma tau}

The \textsc{kosma}-$\tau$ PDR model was created by \citet{Sternberg89} and has since been updated multiple times (e.g., \citealt{Rollig06, Rollig13Szczerba}). The code calculates the emission from a spherical molecular cloud with a radial, power-law density profile. The line ratios from a large grid of model runs, all with a power-law index of 1.5 and a core radius of 0.2 times the total cloud radius, have been made publicly available online. Unfortunately, only the ratios of the lines, and not the absolute line intensities, are made available. All of these model runs have a turbulent doppler parameter of 0.72 km s$^{-1}$, corresponding to an FWHM of $\sim1.2$ km s$^{-1}$.

Only models with Solar metallicities and incident radiation fields equal to the mean ISRF are selected, such that a grid of models that vary only in total mass and the density at the edge of the cloud is obtained. The radius of a model cloud, in cm, is given by $R = 5.3 \times 10^{18} \sqrt[3]{M / n}$, where M is the total cloud mass in Solar masses and n is the density at the outer radius in cm$^{-3}$ \citep{Rollig13Ossenkopf}. The maximum column density, in cm$^{-2}$, of a model is $N = 2 \times 4.7 n R$, where the factor of two accounts for the material on both the front and back side of the sphere \citep{Rollig13Ossenkopf}. Models with densities between 10$^3$ and 10$^5$ cm$^{-3}$ at the outer radius and a column density of at least 10$^{22}$ H nuclei cm$^{-3}$ are selected. The peak column density of B1-E5 is N(H$_2$) = $1.45 \times 10^{22}$ cm$^{-3}$ \citep{Sadavoy12}, a factor of 3 larger than the minimum column density limit imposed for the \textsc{kosma}-$\tau$ models.

The density structure adopted by the \textsc{kosma}-$\tau$ PDR model code \citep{Rollig13Ossenkopf} is
\begin{equation}
n(r) = n_0 \left(\frac{r}{R}\right)^{-\alpha},
\label{eqn:rhoprofile}
\end{equation}
for $R_{core} \le r \le R$ and $n(r) = $ constant for radii less than $R_{core}$. The standard parameters for the \textsc{kosma}-$\tau$ model are $\alpha = 1.5$ and $R_{core}=0.2R$, which roughly approximates the structure of a Bonnor-Ebert sphere \citep{Rollig13Ossenkopf}. For such a density structure, the central density is only an order of magnitude denser than the gas at the very edge of the cloud. 

\subsection{Model Fits}
\label{fits}

\begin{figure*}
   \centering
   \includegraphics[width=6.5in]{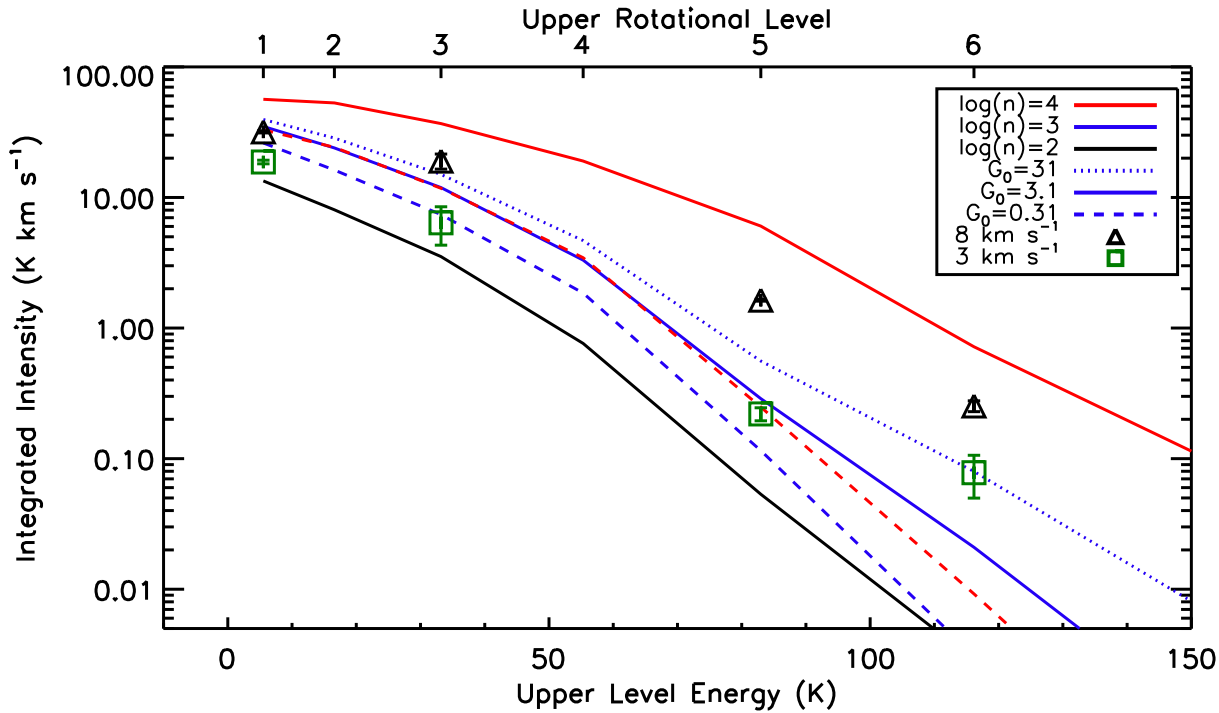}
   \caption{Integrated intensities of various $^{12}$CO lines from the \citet{Kaufman99} PDR models are shown as the lines, while the observed integrated intensities are shown as symbols. Black, blue, and red lines correspond to densities of 10$^2$, 10$^3$, and 10$^4$ cm$^{-3}$, respectively, while dashed, solid, and dotted lines correspond to ISRFs of 0.31, 3.1, and 31 Habing, respectively. Triangles denote integrated intensities of the 8 km s$^{-1}$ component while squares are used for the 3 km s$^{-1}$ component. The error bars show three times the uncertainties in the integrated intensities. The column density for the PDR models has been set to $7 \times 10^{21}$ cm$^{-2}$, to match the total column density of B1-E5.}
   \label{fig:kaufmanint}
\end{figure*}

\begin{figure*}
   \centering
   \includegraphics[width=6.5in]{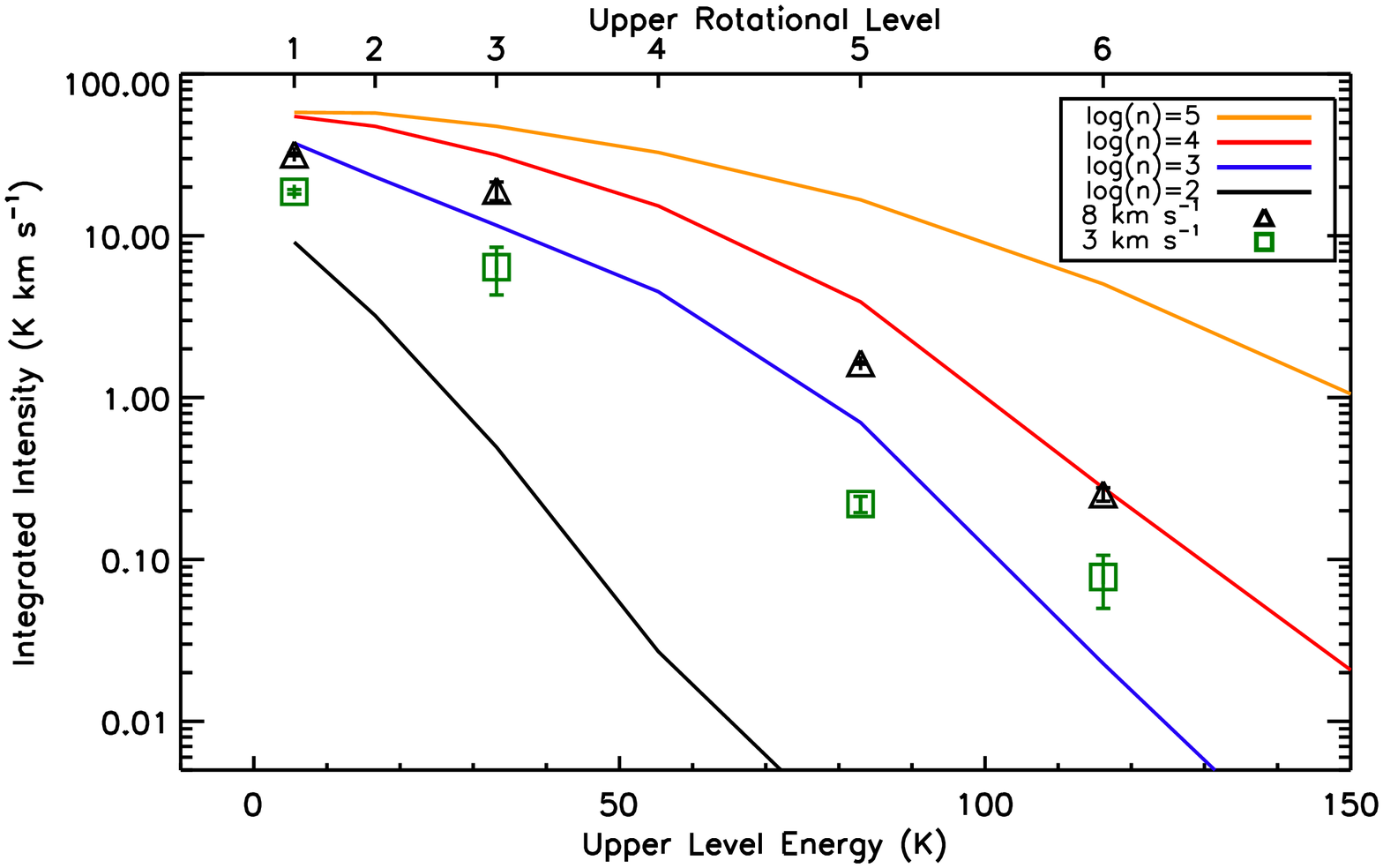}
   \caption{Integrated intensities of various $^{12}$CO lines from the \textsc{meudon} PDR models are shown as the lines, while the observed integrated intensities are shown as symbols. Black, blue, red and gold lines correspond to densities of 10$^2$, 10$^3$, 10$^4$, and 10$^5$ cm$^{-3}$, respectively. Triangles denote integrated intensities of the 8 km s$^{-1}$ component while squares are used for the 3 km s$^{-1}$ component. The error bars show three times the uncertainties in the integrated intensities.}
   \label{fig:meudonint}
\end{figure*}

\begin{figure*}
   \centering
   \includegraphics[width=6.5in]{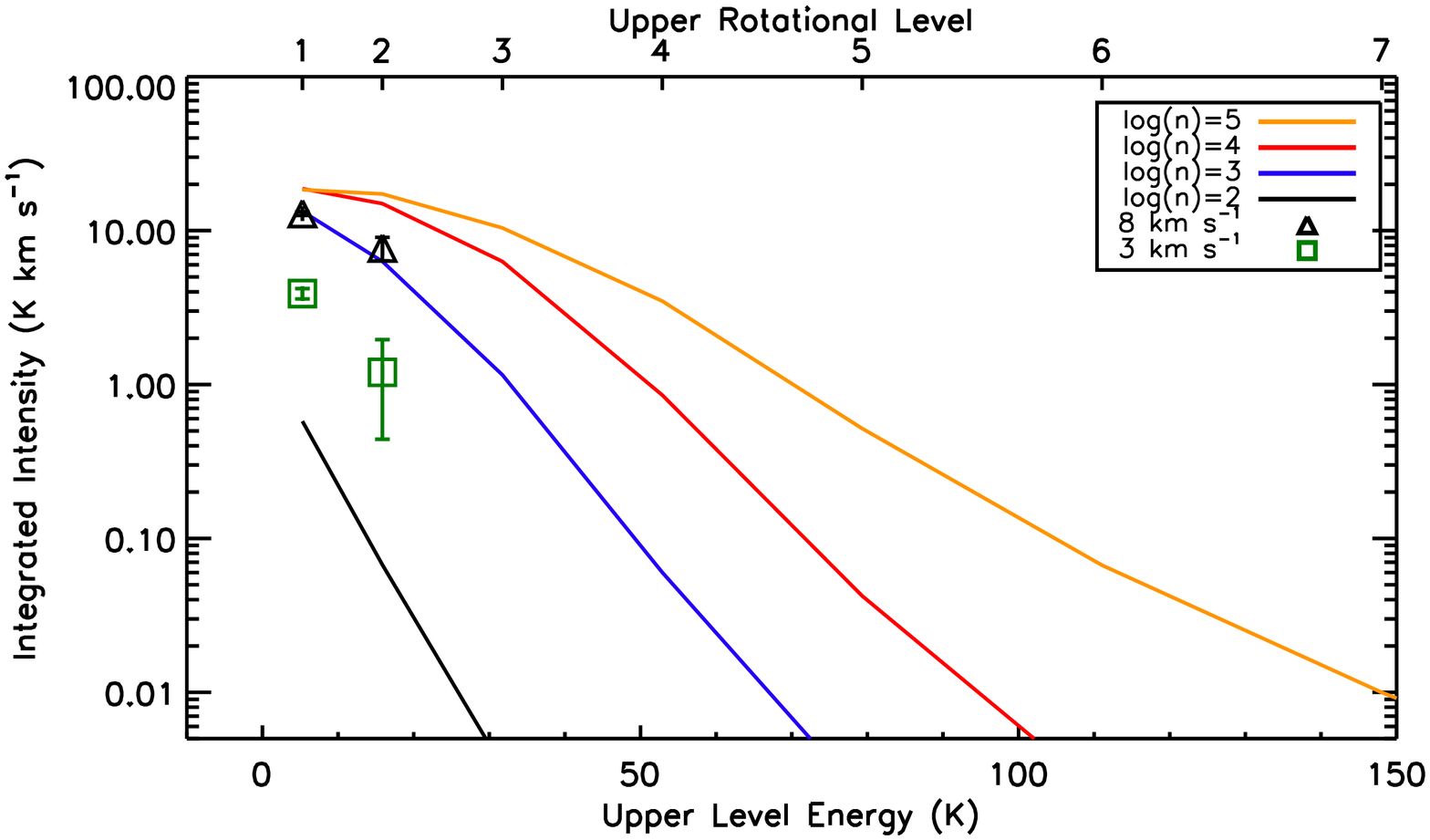}
   \caption{Integrated intensities of various $^{13}$CO lines from the \textsc{meudon} PDR models are shown as the lines, while the observed integrated intensities are shown as symbols. Black, blue, red and gold lines correspond to densities of 10$^2$, 10$^3$, 10$^4$, and 10$^5$ cm$^{-3}$, respectively. Triangles denote integrated intensities of the 8 km s$^{-1}$ component while squares are used for the 3 km s$^{-1}$ component. The error bars show three times the uncertainties in the integrated intensities.}
   \label{fig:meudonint13}
\end{figure*}

Figs \ref{fig:kaufmanint} and \ref{fig:meudonint} show the $^{12}$CO integrated intensities predicted by the \citet{Kaufman99} PDR models and the \textsc{meudon} models, respectively. Fig. \ref{fig:meudonint13} shows the $^{13}$CO integrated intensities predicted by the \textsc{meudon} models. These figures also show the observed integrated intensities towards B1-E5. The error bars show three times the uncertainties in the integrated intensities.

\begin{figure*}
   \centering
   \includegraphics[width=6.5in]{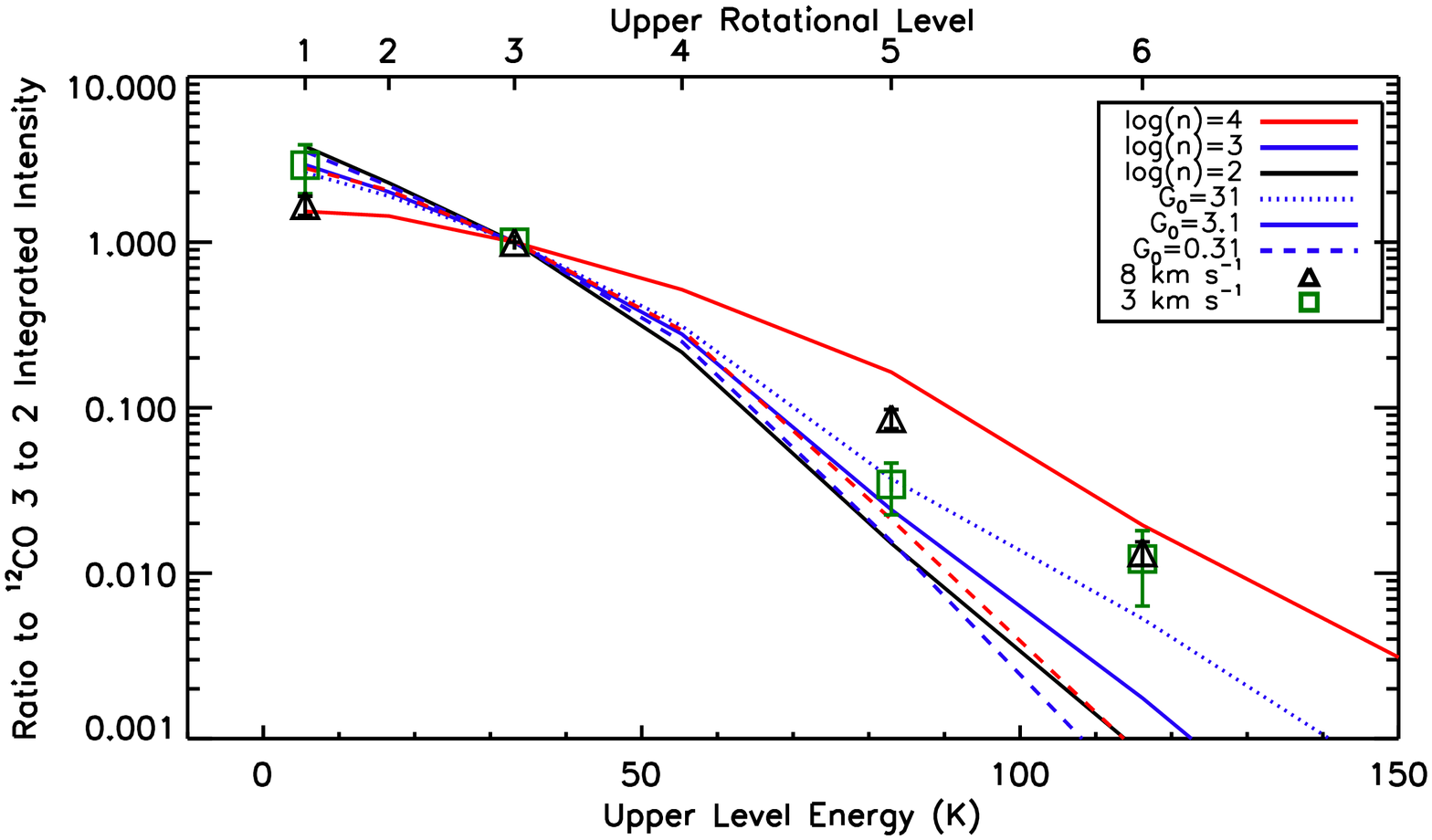}
   \caption{Ratios of the integrated intensities of various $^{12}$CO lines with respect to the $^{12}$CO 3 $\rightarrow$ 2 line. The \citet{Kaufman99} PDR models are shown as lines, while the observed data are shown as symbols. The black, blue, and red lines are for densities of 10$^2$, 10$^3$, and 10$^4$ cm$^{-3}$, respectively, and the dashed, solid, and dotted lines are for ISRFs of 0.31, 3.1, and 31 Habing, respectively. The data for the 3 km s$^{-1}$ component are shown as squares while the 8 km s$^{-1}$ component data are shown as triangles. The error bars show three times the uncertainties in the ratios of the integrated intensities. The column density for the PDR models has been set to $7 \times 10^{21}$ cm$^{-2}$, to match the total column density of B1-E5.}
   \label{fig:kaufmanratio}
\end{figure*}

\begin{figure*}
   \centering
   \includegraphics[width=6.5in]{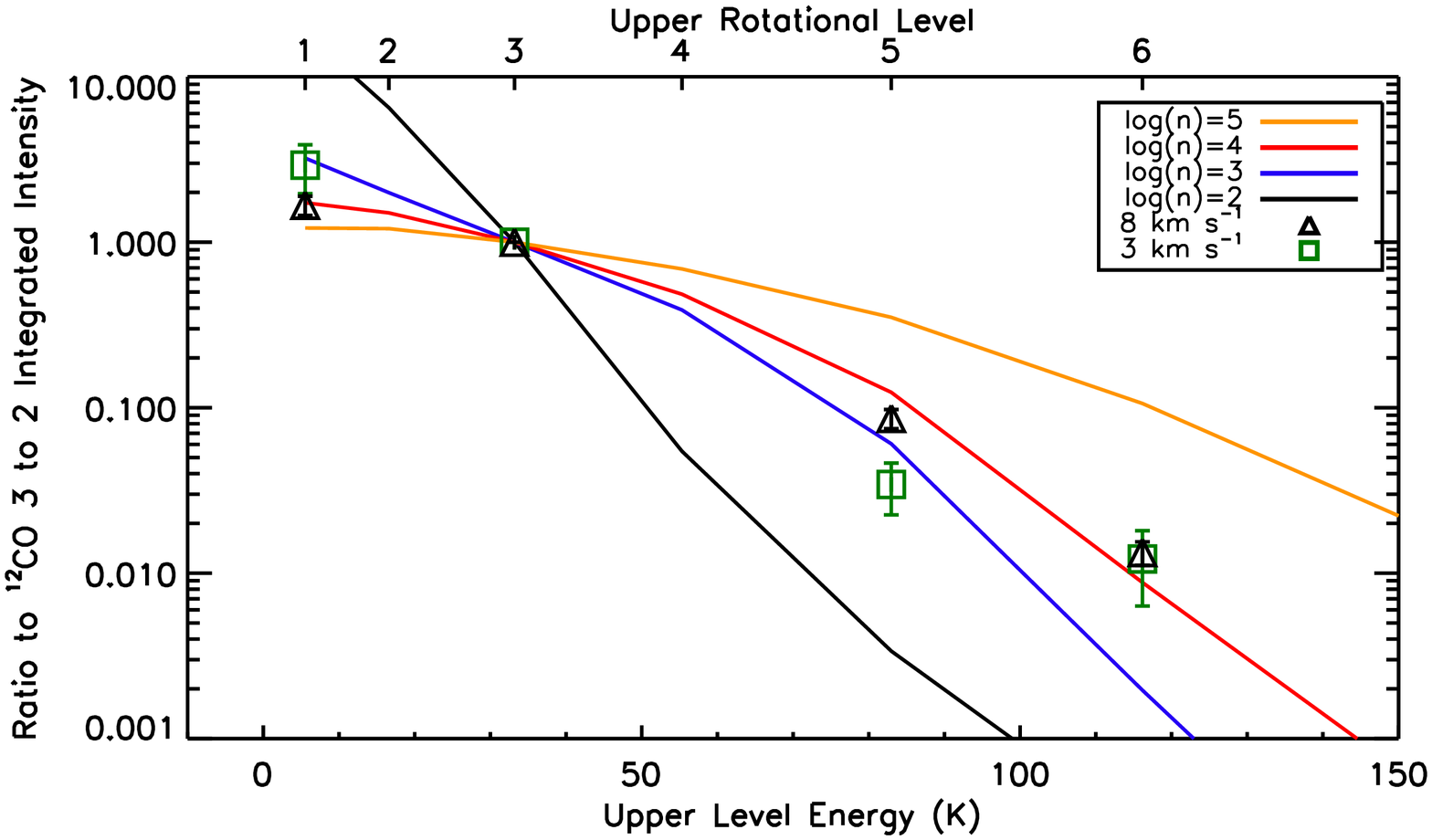}
   \caption{Ratios of the integrated intensities of various $^{12}$CO lines with respect to the $^{12}$CO 3 $\rightarrow$ 2 line. The \textsc{meudon} PDR models are shown as lines, while the observed data are shown as symbols. The black, blue, red, and gold lines are for densities of 10$^2$, 10$^3$, 10$^4$, and 10$^5$ cm$^{-3}$, respectively. The data for the 3 km s$^{-1}$ component are shown as squares while the 8 km s$^{-1}$ component data are shown as triangles. The error bars show three times the uncertainties in the ratios of the integrated intensities. The error bar associated with the highest upper level energy point that the log(n) = 4 model goes through is the larger error bar associated with the 3 km s$^{-1}$ component. The 8 km s$^{-1}$ error range lies significantly above the log(n) = 4 curve for this point.}
   \label{fig:meudonratio}
\end{figure*}

\begin{figure*}
   \centering
   \includegraphics[width=6.5in]{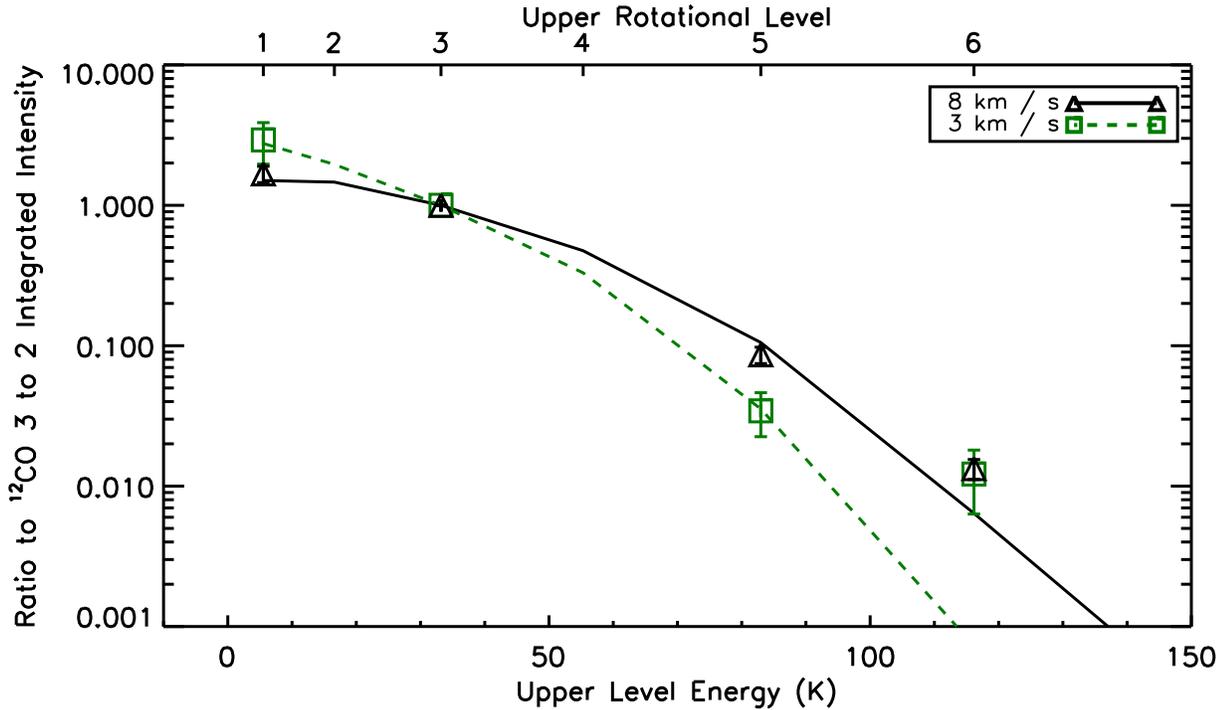}
   \caption{Ratios of the integrated intensities of various $^{12}$CO lines with respect to the $^{12}$CO 3 $\rightarrow$ 2 line. The solid and dashed lines are the best fitting \textsc{kosma}-$\tau$ models for the 8 and 3 km s$^{-1}$ components, respectively. The observed ratios of the 8 and 3 km s$^{-1}$ components are shown as the triangles and squares, respectively. The error bars show three times the uncertainties in the ratios of the integrated intensities and are the same as in Fig. \ref{fig:meudonratio}. As such, the error bar associated with the highest upper level energy point that the 8 km s$^{-1}$ model goes through is the larger error bar associated with the 3 km s$^{-1}$ component.}
   \label{fig:kosmaratio}
\end{figure*}

\begin{figure*}
   \centering
   \includegraphics[width=6.5in]{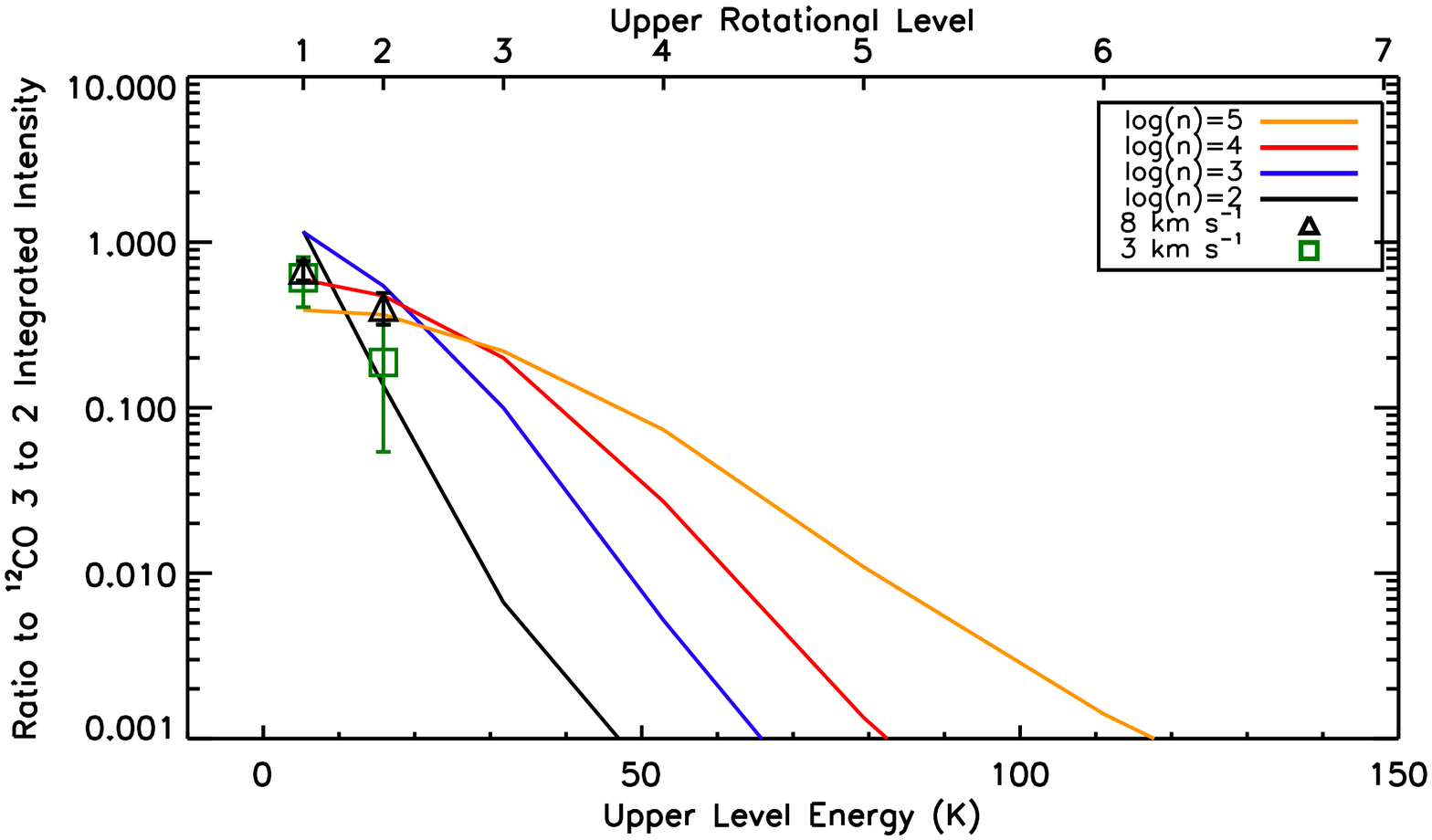}
   \caption{Ratios of the integrated intensities of various $^{13}$CO lines with respect to the $^{12}$CO 3 $\rightarrow$ 2 line. The \textsc{meudon} PDR models are shown as lines, while the observed data are shown as symbols. The black, blue, red, and gold lines are for densities of 10$^2$, 10$^3$, 10$^4$, and 10$^5$ cm$^{-3}$, respectively. The data for the 3 km s$^{-1}$ component are shown as squares while the 8 km s$^{-1}$ component data are shown as triangles. The error bars show three times the uncertainties in the ratios of the integrated intensities.}
   \label{fig:meudonratio13}
\end{figure*}

It is reasonably common (e.g., \citealt{Wolfire90,Lord96}) to fit the ratio of observed lines, rather than the integrated intensities themselves, in case of non-unity beam filling factors. Figs \ref{fig:kaufmanratio}-\ref{fig:kosmaratio} show the integrated intensity ratios of various $^{12}$CO lines with respect to the $^{12}$CO J = 3 $\rightarrow$ 2 line, as predicted by the \citet{Kaufman99} PDR models, the \textsc{meudon} models, and the \textsc{kosma}-$\tau$ models, respectively. Fig. \ref{fig:meudonratio13} shows the ratios of various $^{13}$CO lines to the $^{12}$CO J = 3 $\rightarrow$ 2 line, as predicted by the \textsc{meudon} models. The ratios of the observed line integrated intensities to the observed integrated intensity of the $^{12}$CO J = 3 $\rightarrow$ 2 line are also shown. The error bars show three times the uncertainties in the ratios.

For the \textsc{kosma}-$\tau$ models, we perform a chi-square fitting of the ratios of the $^{12}$CO 6 $\rightarrow$ 5, 5 $\rightarrow$ 4, and 1 $\rightarrow$ 0 lines to the $^{12}$CO 3 $\rightarrow$ 2 line, weighted appropriately for the estimated integrated intensity uncertainties, and we only show the best-fitting models for the 3 and 8 km s$^{-1}$ components in Fig. \ref{fig:kosmaratio}. For the 3 km s$^{-1}$ component, the best fit is provided by a model with a density of 10$^{4.2}$ cm$^{-3}$ and column density of $4 \times 10^{22}$ cm$^{-2}$, while for the 8 km s$^{-1}$ component, the best-fitting model has a density just under 10$^5$ cm$^{-3}$ and a column density of $1 \times 10^{22}$ cm$^{-2}$. 

In general, since larger ISRFs imply a larger incoming energy flux, the integrated intensities of the CO lines increase with increasing $G_0$, in order for a cloud to maintain an equilibrium between heating and cooling terms. The CO SLED also becomes flatter with increasing $G_0$, as the temperature of CO emitting gas increases. This latter effect is not necessarily trivially expected, as a larger ISRF will cause CO, and other molecules, to only form at larger optical depths, where the gas need not be significantly warmer than where CO forms in lower ISRFs. Higher densities also produce larger intensities and flatter SLEDs, as the gas density approaches the large critical densities of the higher J lines. 

Based upon the integrated intensities, the best-fitting models with an order of unity ISRF have a density of 10$^3$ cm$^{-3}$, for both the \citet{Kaufman99} and \textsc{meudon} codes. These models do a reasonable job at reproducing the integrated intensities of lower J lines, both $^{13}$CO and $^{12}$CO lines for the \textsc{meudon} code, but underpredict the integrated intensity of the $^{12}$CO J = 6 $\rightarrow$ 5 line. A lower $G_0$ of 0.31 and a higher density of 10$^4$ cm$^{-3}$ also produces reasonable integrated intensities from the \citet{Kaufman99} code. Since the 3 km s$^{-1}$ component has lower integrated intensities, these models suggest that the 3 km s$^{-1}$ component traces lower density gas than the 8 km s$^{-1}$ component. This is consistent with the 3 km s$^{-1}$ component having lower integrated intensities in C$^{18}$O and NH$_3$ lines (\citealt{Sadavoy12}; Sadavoy et al., in preparation). 

For the 3 km s$^{-1}$ component, the \citet{Kaufman99} models with a density and ISRF of 10$^{3}$ cm$^{-3}$ and 3.1 Habing, or 10$^{4}$ cm$^{-3}$ and 0.31 Habing, also reasonably reproduce the line ratios of the $^{12}$CO J = 1 $\rightarrow$ 0 and 5 $\rightarrow$ 4 lines, with respect to the 3 $\rightarrow$ 2 line. The best-fitting \textsc{meudon} model to the 3 km s$^{-1}$ ratios also has a density of 10$^3$ cm$^{-3}$, but the best-fitting \textsc{kosma}-$\tau$ model has a slightly higher density of 10$^{4.2}$ cm$^{-3}$. The $^{13}$CO ratios are roughly consistent with the \textsc{meudon} models. All three codes underpredict the $^{12}$CO J = 6 $\rightarrow$ 5 to 3 $\rightarrow$ 2 ratio, with the \textsc{kosma}-$\tau$ model being a factor of 17 low, a 6$\sigma$ discrepancy. 

The ratio of the $^{12}$CO J = 1 $\rightarrow$ 0 to 3 $\rightarrow$ 2 lines is much closer to unity for the 8 km s$^{-1}$ component than the 3 km s$^{-1}$ component, which makes higher density models (10$^4$ cm$^{-3}$ for the \citealt{Kaufman99} and \textsc{meudon} codes, 10$^5$ cm$^{-3}$ for the \textsc{kosma}-$\tau$ code) fit this ratio better for the 8 km s$^{-1}$ component than the 3 km s$^{-1}$ component. The \textsc{meudon} 10$^4$ cm$^{-3}$ model reasonably reproduces the integrated intensity ratios of the $^{12}$CO J = 5 $\rightarrow$ 4 and 1 $\rightarrow$ 0 lines with the 3 $\rightarrow$ 2 line, but again slightly underpredicts the ratio of the 6 $\rightarrow$ 5 to 3 $\rightarrow$ 2 lines. In Fig. \ref{fig:meudonratio}, the error bar for the 6 $\rightarrow$ 5 point that crosses the 10$^{4}$ cm$^{-3}$ model line is that of the 3 km s$^{-1}$ component, not the 8 km s$^{-1}$ component. The 8 km s$^{-1}$ component error range for the ratio of the 6 $\rightarrow$ 5 to 3 $\rightarrow$ 2 lines is not consistent with the \textsc{meudon} 10$^{4}$ cm$^{-3}$ model. Similarly, while the best-fitting \textsc{kosma}-$\tau$ model to the 8 km s$^{-1}$ component matches the line ratios of the $^{12}$CO J = 5 $\rightarrow$ 4 and lower lines, it underpredicts the 6 $\rightarrow$ 5 to 3 $\rightarrow$ 2 line ratio by a factor of 2, which is a 9$\sigma$ discrepancy due to the high signal-to-noise ratio of the 8 km s$^{-1}$ component data. 

For the 8 km s$^{-1}$ component, the \citet{Kaufman99} 10$^{4}$  cm$^{-3}$ model overpredicts the $^{12}$CO J = 5 $\rightarrow$ 4 and 6 $\rightarrow$ 5 ratios with the 3 $\rightarrow$ 2 line while matching the 1 $\rightarrow$ 0 to 3 $\rightarrow$ 2 ratio. Alternatively, for the 8 km s$^{-1}$ component, a lower density \citet{Kaufman99} model (10$^3$ cm$^{-3}$) can be found that fits the ratio of the $^{12}$CO J = 5 $\rightarrow$ 4 to 3 $\rightarrow$ 2 lines, but overpredicts the 1$\rightarrow$ 0 line and underpredicts the 6 $\rightarrow$ 5 line, relative to the 3 $\rightarrow$ 2 line. From the line ratios alone, it is not obvious whether the higher density model, which matches the 1 $\rightarrow$ 0 to 3 $\rightarrow$ 2 ratio, or the lower density model, which matches the 5 $\rightarrow$ 4 to 3 $\rightarrow$ 2 ratio, provides the better fit to the data. Since the lower density model, however, does a better job at predicting the absolute integrated intensities of the lines, we prefer the lower density solution to the 8 km s$^{-1}$ line ratios for the \citet{Kaufman99} code. 

Overall, any PDR model with a reasonable density, column density, and ISRF for B1-E5 that matches the observed integrated intensities of the $^{12}$CO J = 3 $\rightarrow$ 2 and lower lines underpredicts the integrated intensity of the $^{12}$CO J = 6 $\rightarrow$ 5 line, for either component. Similarly, any model that matches the $^{12}$CO J = 5 $\rightarrow$ 4 to 3 $\rightarrow$ 2 ratio significantly underpredicts the $^{12}$CO J = 6 $\rightarrow$ 5 to 3 $\rightarrow$ 2 integrated intensity ratio. 

\begin{figure*}
   \centering
   \includegraphics[width=6.5in]{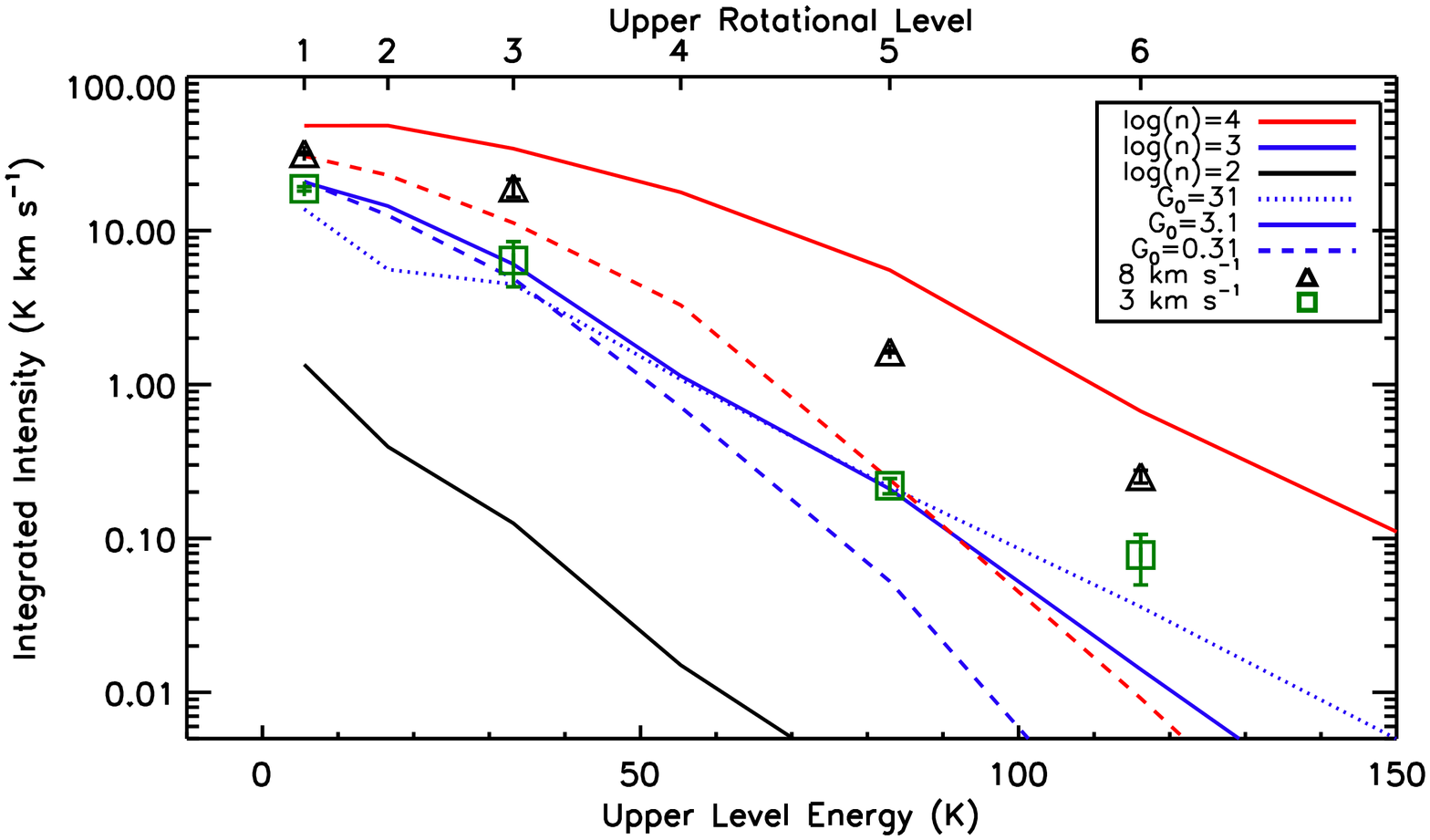}
   \caption{Integrated intensities of various $^{12}$CO lines from the \citet{Kaufman99} PDR models are shown as the lines, while the observed integrated intensities are shown as symbols. Black, blue, and red lines correspond to densities of 10$^2$, 10$^3$, and 10$^4$ cm$^{-3}$, respectively, while dashed, solid, and dotted lines correspond to ISRFs of 0.31, 3.1, and 31 Habing, respectively. Triangles denote integrated intensities of the 8 km s$^{-1}$ component while squares are used for the 3 km s$^{-1}$ component. The error bars show three times the uncertainties in the integrated intensities. The column density for the PDR models has been set to $2 \times 10^{21}$ cm$^{-2}$.}
   \label{fig:kaufmanint3kms}
\end{figure*}

\begin{figure*}
   \centering
   \includegraphics[width=6.5in]{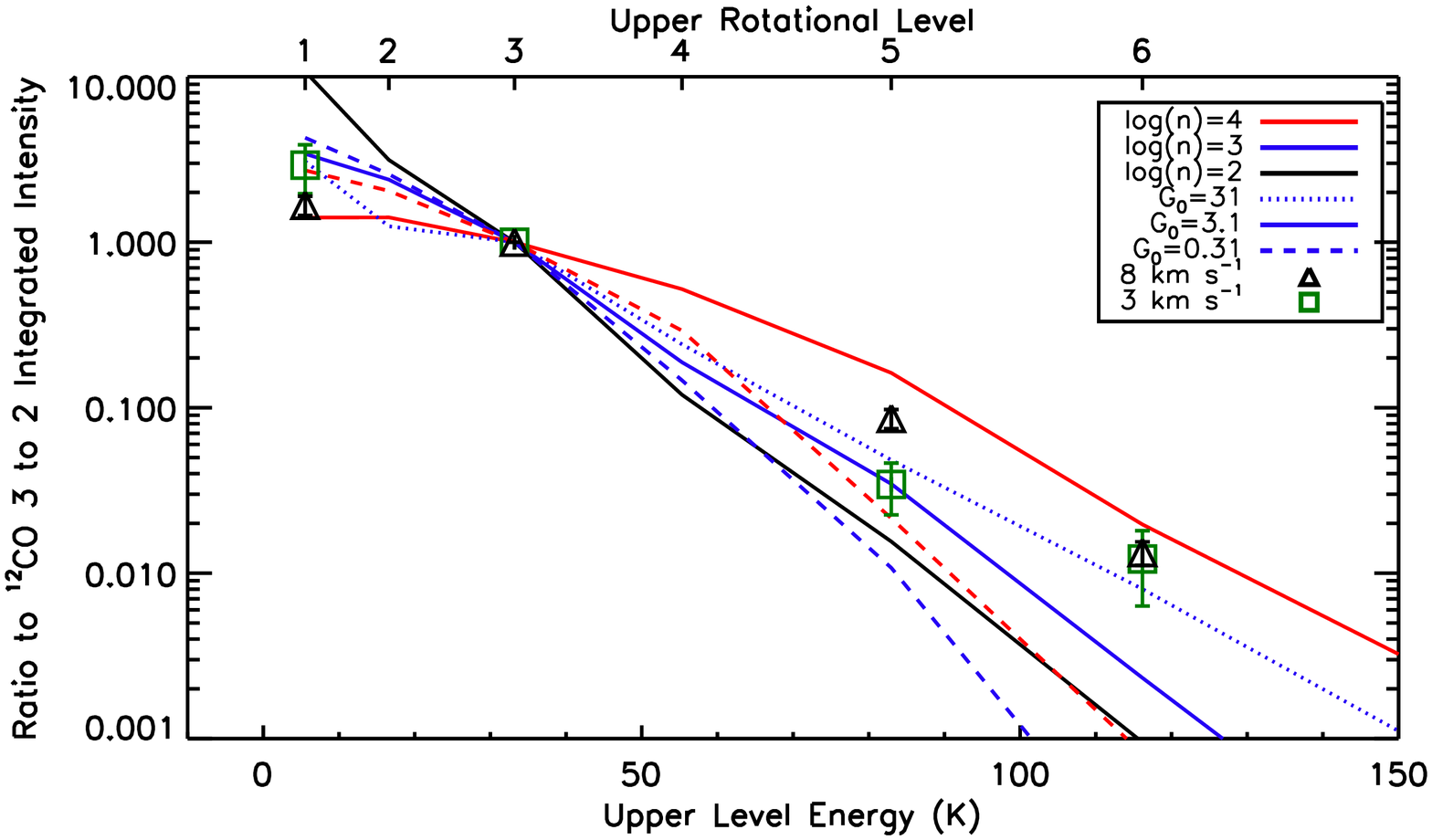}
   \caption{Ratios of the integrated intensities of various $^{12}$CO lines with respect to the $^{12}$CO 3 $\rightarrow$ 2 line. The \citet{Kaufman99} PDR models are shown as lines, while the observed data are shown as symbols. The black, blue, and red lines are for densities of 10$^2$, 10$^3$, and 10$^4$ cm$^{-3}$, respectively, and the dashed, solid, and dotted lines are for ISRFs of 0.31, 3.1, and 31 Habing, respectively. The data for the 3 km s$^{-1}$ component are shown as squares while the 8 km s$^{-1}$ component data are shown as triangles. The error bars show three times the uncertainties in the ratios of the integrated intensities. The column density for the PDR models has been set to $2 \times 10^{21}$ cm$^{-2}$.}
   \label{fig:kaufmanratio3kms}
\end{figure*}

Equation \ref{eqn:column} suggests that one reason the 3 km s$^{-1}$ component may be dimmer than the 8 km s$^{-1}$ component is that the 3 km s$^{-1}$ component may come from gas with a lower column density than the entirety of B1-E5. Figs \ref{fig:kaufmanint3kms} and \ref{fig:kaufmanratio3kms} show the \citet{Kaufman99} PDR models when a lower column density of $2 \times 10^{21}$ cm$^{-2}$ is used. Reducing the column density tends to preferentially reduce the integrated intensities of the lower J lines, but, as with the higher column density models, any model that fits the $^{12}$CO J = 1 $\rightarrow$ 0, 3 $\rightarrow$ 2, and 5 $\rightarrow$ 4 lines underpredicts the 6 $\rightarrow$ 5 line. 

\section{DISCUSSION}
\label{discussion}

All three sets of PDR models that have been fit to our data, the \citet{Kaufman99} models, the \textsc{meudon} models, and the \textsc{kosma}-$\tau$ models, fail to simultaneously explain the ratios of the observed integrated intensities of the $^{12}$CO J = 1 $\rightarrow$ 0, 3 $\rightarrow$ 2, 5 $\rightarrow$ 4, and 6 $\rightarrow$ 5 lines. All three sets of models underpredict the $^{12}$CO J = 6 $\rightarrow$ 5 line when the lower three lines are reasonably well fit. One possible interpretation of these observations is that there is a warm gas component of B1-E5 that is unaccounted for in the PDR models. 

\subsection{Comparison to Shock Models}
\label{shock}

Turbulent energy dissipation is one of the primary heating sources for deeply embedded molecular gas with densities similar to B1-E5 (e.g., \citealt{Glover12a, Pon12Kaufman}) and this turbulent heating is not included within any of the three PDR codes used. While the turbulent energy dissipation rate is typically of the order of the cosmic ray heating rate (e.g., \citealt{Glover12a, Pon12Kaufman}), the turbulent energy is not necessarily dissipated over the entire volume of a cloud. If the turbulent energy is dissipated over a small fraction of the cloud, such as in localized shock fronts, then this turbulent energy dissipation could produce a small, warm gas component in a molecular cloud that would lead to enhanced emission in higher J CO lines. 

\citet{Pon12Kaufman} estimate the CO integrated intensity that would come from a GMC, assuming that turbulence decays on the sound crossing time of the cloud and that all of this turbulent energy is dissipated in low-velocity shocks (2 or 3 km s$^{-1}$) generated from the turbulent motions. They predict that shocks should generate warm gas, between 50 and 150 K, that would only have a volume filling factor of a few tenths of a percent. This small volume of shocked gas would be expected to be distributed relatively uniformly across the entire molecular cloud, as turbulent shocks should be forming throughout a cloud. Despite this low volume filling factor, such a warm gas component would produce more emission in the CO J = 6 $\rightarrow$ 5 transition than the rest of the molecular cloud. As such, the excess $^{12}$CO J = 6 $\rightarrow$ 5 emission observed towards B1-E5 may be due to emission from low-velocity shocks within B1-E5. 

\citet{Pon12Kaufman} examined shock models with pre-shock molecular H$_2$ densities between 10$^{2.5}$ and 10$^{3.5}$ cm$^{-3}$ and shock velocities of 2 and 3 km s$^{-1}$. Since the best-fitting PDR models for B1-E5 generally have densities between 10$^3$ and 10$^4$ cm$^{-3}$, consistent with the average density of 10$^4$ cm$^{-3}$ derived by \citet{Sadavoy12}, we chose to compare our observations to the higher density (10$^{3.5}$ cm$^{-3}$) shock models of \citet{Pon12Kaufman}. Based on the average FWHM of 1.8 km s $^{-1}$ for the WBS observations, we further select a shock model with a shock velocity of 2 km s$^{-1}$. Thus, we select model n35v2b1 from \citet{Pon12Kaufman} to compare against our observations. 

\begin{figure*}
   \centering
   \includegraphics[width=6.5in]{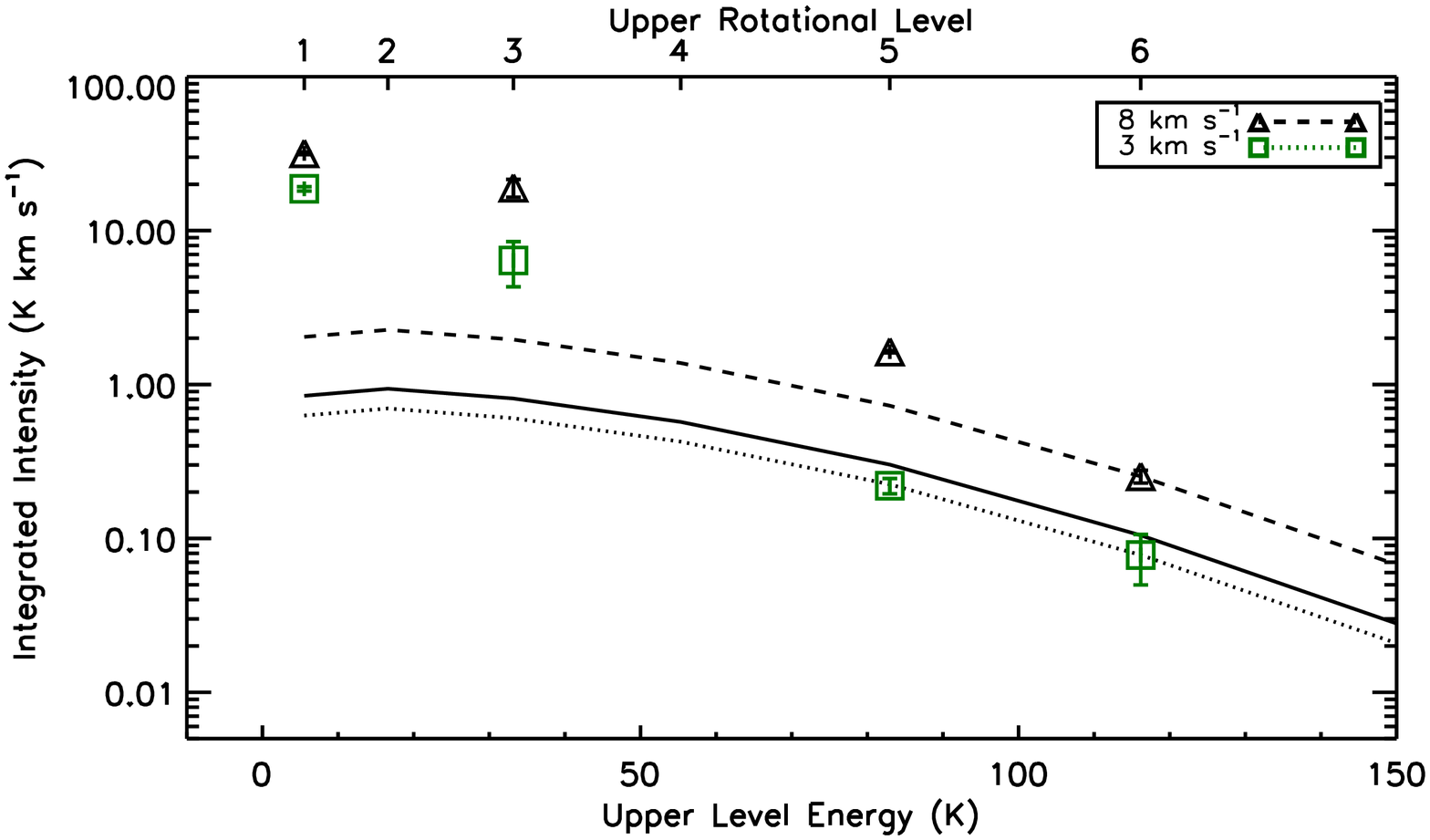}
   \caption{Integrated intensities of various $^{12}$CO lines as predicted by the n35v2b1 shock model of \citet{Pon12Kaufman}. This is a shock model with an H$_2$ density of 10$^{3.5}$ cm$^{-3}$ and a shock velocity of 2 km s$^{-1}$. The solid line gives the values predicted by \citet{Pon12Kaufman} while the dashed and dotted lines show the same model scaled by factors of 2.4 and 0.74, respectively, to match the observed integrated intensity of the 6 $\rightarrow$ 5 transition of the 8 and 3 km s$^{-1}$ components, respectively. Such scalings are equivalent to changing the volume filling factor of the shocked gas. The observed integrated intensities are shown as the symbols, with the data for the 3 km s$^{-1}$ component being shown as squares and the 8 km s$^{-1}$ component data being shown as triangles. The error bars show three times the uncertainties in the integrated intensities.}
   \label{fig:shockmodels}
\end{figure*}

Fig. \ref{fig:shockmodels} shows the $^{12}$CO integrated intensities predicted by the n35v2b1 shock model. It also shows the same shock models scaled by factors of 0.7 and 2.4 to match the observed $^{12}$CO J = 6 $\rightarrow$ 5 integrated intensities of the 3 and 8 km s$^{-1}$ components, respectively. Such a scaling of the integrated intensities is equivalent to scaling the volume filling factor of the shocks. For both the 3 and 8 km s$^{-1}$ components, the shock model can reasonably explain the high $^{12}$CO J = 6 $\rightarrow$ 5 integrated intensity without overpredicting any of the lower J lines. Underpredictions of the lower J lines are acceptable as the lower J lines are expected to have a significant PDR component \citep{Pon12Kaufman}. 

To derive a volume filling factor of the shock-heated gas, the radius of a cloud must be known. \citet{Pon12Kaufman} make the assumption that clouds obey the velocity--size relation of Larson's law \citep{Larson81,Solomon87} and, with the knowledge that the cooling length for model n35v2b1 is 0.01 pc, find that clouds should have a volume filling factor of shocked gas of 0.07 per cent. The shock cooling length of 0.01 pc is of the order of the size of the transition zone from turbulent motions to coherent motions observed in the B5 core \citep{Pineda10} and is of the order of the size of B1-E5. 

The observed integrated intensity of the $^{12}$CO J = 6 $\rightarrow$ 5 line can also be used to estimate the volume filling factor of the shocked gas. The total CO luminosity can be determined from the observed integrated intensity if the radius of the emitting region is known. Since the shock model calculates the energy radiated per unit area of shock front and the cooling length of the shocked gas, the CO luminosity can then be used to calculate the volume of shocked gas within B1-E5. The volume filling factor of the shocked gas then comes about readily by dividing by the total volume of the emitting region. We assume that the emitting region in B1-E is a spherical region with a radius of 1 pc, roughly the size of the B1-E region. The average FWHM from the WBS data, 1.8 km s$^{-1}$, when used with the velocity--size relation from \citet{Solomon87}:
\begin{equation}
\sigma = 0.72 (R / \mbox{pc})^{0.5} \mbox{km s}^{-1}
\end{equation}
suggests a similar radius of 1.1 pc. With this 1 pc radius, the observed integrated intensities of the 6 $\rightarrow$ 5 line can be explained if the volume filling factors of the shock-heated gas are 0.04 and 0.11 per cent for the 3 and 8 km s$^{-1}$ components, respectively. The total volume filling factor of the shock-heated gas is thus 0.15 per cent. 

Since the total volume of shocked gas can be calculated from the observed integrated intensity of the $^{12}$CO 6 $\rightarrow$ 5 line, the total rate of  turbulent energy dissipation via low-velocity shocks in B1-E can also be calculated. Accounting for the emission from both the 3 and 8 km s$^{-1}$ components, the total turbulent energy dissipation rate in B1-E5 is thus $3.5 \times 10^{32}$ erg s$^{-1}$. 

As shown by \citet{Basu01}, the turbulent energy dissipation rate, $L_{turb}$ can be written as
\begin{equation}
L_{turb} = \frac{\pi \rho \sigma^3 R^2}{\kappa},
\end{equation}
where $\rho$ is the gas density, $\sigma$ is the 1D velocity dispersion of the gas, $R$ is the radius of the cloud, and $\kappa$ is the ratio between the dissipation time and the flow crossing time, which is given by $t_c = 2R / \sigma$. Taking the mean number density of H$_2$ molecules to be 10$^{3.5}$ cm$^{-3}$; the mean mass per hydrogen molecule to be $4.6 \times 10^{-24}$ g, or about 2.77 amu \citep{Kaufman96I};  and the 1D velocity dispersion as 0.6 km s$^{-1}$, based upon a shock velocity of 2 km s$^{-1}$ and a factor of 3.2 scaling between the 1D velocity dispersion and the characteristic shock velocity \citep{Pon12Kaufman}, we find that $\kappa$ is roughly 1/3. This value is independent of the radius adopted.

The linewidths of the $^{12}$CO J = 6 $\rightarrow$ 5 and 5 $\rightarrow$ 4 transitions are of the order of, if not slightly smaller than, the linewidths of the lower J CO lines. This consistency in linewidths does not exclude a low-velocity shock origin for the higher J lines. For shock velocities of 2 or 3 km s$^{-1}$, as modelled by \citet{Pon12Kaufman}, the maximum temperature reached is of the order of 120 K, which would produce a thermal linewidth of only 0.4 km s$^{-1}$, still much smaller than the turbulent velocity dispersion of the gas. These shock velocities are also consistent with the velocity dispersion of B1-E5, as inferred from the low-J lines. 

Observations of the $^{12}$CO J = 7 $\rightarrow$ 6 transition towards B1-E5 would provide a clear confirmation of whether the mid-J CO integrated intensities are larger than predicted from PDR models and the ratio of the 7 $\rightarrow$ 6 to 6 $\rightarrow$ 5 lines would allow for the verification of the typical temperature of the shocked gas, and thus the typical shock strength within B1-E5.

Excess emission in higher J CO lines should not be limited to the B1-E5 region, as it should be a common property of any region with properties similar to B1-E (e.g., low density, low ISRF, etc.). Since Perseus is a northern sky source and difficult to observe with the Atacama Large Millimeter Array (ALMA), it would be very interesting to find a similar source in the Southern hemisphere that could be followed up with ALMA. The high resolution and sensitivity of ALMA should make it possible to characterize the small-scale structure of the mid-J emitting gas and, if the emission is from shock fronts, ALMA should be capable of resolving and thus studying individual shock fronts. The spatial structure of this emission should help discriminate between different possible sources of this mid-J emission. Based on the shock models of \citet{Pon12Kaufman}, if the gas heated from a face-on shock fully filled the ALMA beam, and had a depth equal to the cooling length (0.01 pc), the CO 6 $\rightarrow$ 5 emission would have an integrated intensity of approximately 1 K km s$^{-1}$. If a shock front were instead observed edge-on, such that the depth of the shock emission were 0.05 pc (similar to the length-scale of the transition from turbulent to coherent gas motions observed by \citealt{Pineda10} around the B5 dense core) and the shocked emission filled the ALMA beam, the CO 6 $\rightarrow$ 5 would have an integrated intensity of $\sim5$ K km s$^{-1}$. 

\subsection{Alternative Explanations}
\label{Alternatives}

Other potential sources of a warm gas component include ion-neutral friction \citep{Hennebelle13Andre}, magnetic vortices (e.g. \citealt{Falgarone09, Godard09, Falgarone10Ossenkopf}), and feedback from young stars and protostars, although magnetic vortices are not expected to function at the higher densities of molecular clouds and B1-E5 is reasonably distant from the nearest protostar in the Perseus molecular cloud (e.g., \citealt{Evans09}). Outflows from protostars have velocities of the order of tens of kilometres per second and can generate temperatures in excess of 1000 K (e.g., \citealt{Arce07}). Such large temperatures and strong shocks would generate an even greater enhancement of the mid-J CO lines than observed (e.g., \citealt{Kaufman96II, Lesaffre13}), due to the very large temperatures generated by the outflows. 

Work on the handling of dust properties, including the inclusion of polycyclic aromatic hydrocarbons and H$_2$ formation rates, is currently ongoing in the PDR community and changes in these dust properties can lead to significant changes in the temperature profile of a PDR (e.g., \citealt{Hollenbach12,Rollig13Szczerba}). In turn, this can lead to significant changes in the integrated intensities of higher J lines if more warm gas is produced in the periphery of a cloud, particularly in clumpier clouds \citep{Rollig13Szczerba}. 

Another possible heating source that could produce a warm gas component within B1-E5 would be shock heating due to large-scale flows in the vicinity of B1-E5. While this is similar to the shock heating scenario put forth by \citet{Pon12Kaufman}, with the warm gas component being generated from low-velocity shocks, the origin of these shocks in a large-scale flow is very different from the picture of \citet{Pon12Kaufman} where the shocks come from the generic, widely spread turbulence of a molecular cloud. The northeastern half of the Perseus molecular cloud is known to be interacting with a large bubble produced by a B star \citep{Ridge06Schnee, Pineda08}. Such an expanding bubble could generate shocks within the Perseus molecular cloud. The edge of this bubble, however, is believed to be slightly to the north-east of B1-E, such that the bubble is likely not yet interacting with the B1-E region. This bubble is labelled as CPS 5 in \citet{Arce11}. \citet{Arce11} also point out a smaller bubble, CPS 4, which B1-E lies on the eastern edge of, that might create shocks within B1-E. Such a large-scale interaction with a bubble might, however, produce reasonably large volume filling factors for the shock-heated gas. Similarly, if B1-E is currently accreting material from the surrounding cloud, this accretion could generate low-velocity shocks, although there are no data currently indicating such an infall.

These observations were not taken with the same beam sizes and we have been unable to smooth the {\it Herschel} data to larger beam sizes, since we did not make a map of the B1-E region. As such, variations in the beam filling factor may also alter the observed ratios. Non-unity beam filling factors will only lower the observed intensity in a line, such that non-unity beam filling factors could only plausibly work for the higher density PDR models. Assuming that the emitting region in B1-E5 is centrally concentrated in the beams, smaller beams should have larger beam filling factors, such that observations taken with smaller beams should be closer to the predicted PDR fluxes. For the highest density \citet{Kaufman99} and \textsc{meudon} models, the higher J lines would require smaller filling factors than the low-J lines, despite the higher J lines having been observed with smaller beams. All of the other models either underpredict some of the lines or require significantly different volume filling factors for the $^{12}$CO 5 $\rightarrow$ 4 and 6 $\rightarrow$ 5 lines, which is highly improbable given the similarities in beam sizes used to observe these two lines. Furthermore, the Sadavoy et al. (in preparation) data show that the $^{13}$CO J = 2 $\rightarrow$ 1 integrated intensity is roughly the same in a 12 arcsec beam as in a 114 arcsec beam, indicating that the beam filling factor likely remains the same on all size scales probed by the data that we have used. 

Smaller linewidths would create slightly flatter SLEDs, as less energy could be emitted from the optically thick, lower lines. The assumed linewidths in the \textsc{meudon} PDR models, however, is already on the lower end of the observed FWHM, such that changes in the FWHM of the PDR models is highly unlikely to produce models consistent with the 6 $\rightarrow$ 5 integrated intensity. 

Constant density, constant temperature \textsc{radex} models were run for kinetic temperatures between 5 and 20 K, gas densities between 10$^2$ and 10$^5$ cm$^{-3}$, and CO column densities between $7.3 \times 10^{17}$ and $2.9 \times 10^{18}$ cm$^{-2}$. None of the models were able to reproduce the large observed CO 6 $\rightarrow$ 5 integrated intensity of the 3 km s$^{-1}$ component while fitting the lower J lines. For the 8 km s$^{-1}$ component, however, \textsc{radex} models with densities of 10$^3$ to 10$^4$ cm$^{-3}$ and large kinetic temperatures of 20 K are able to fit all of the observed lines reasonably well. Such a large temperature is hard to reconcile with our theoretical understandings of cloud cooling and heating from external radiation fields and cosmic ray ionization (e.g., PDR models). \citet{Sadavoy12} place a 19 K upper limit on the temperature of B1-E5 and find a temperature of approximately 10 K for the B1-E2 structure, the only substructure within B1-E for which they do not only have an upper limit for the temperature.

\section{CONCLUSIONS}
\label{conclusions}

The Perseus B1-E region is in the process of forming prestellar cores and thus, presents an ideal laboratory to study the heating of dense molecular gas by turbulent energy dissipation without the influence of feedback from protostars. We obtained observations of the $^{12}$CO J = 6 $\rightarrow$ 5 and 5 $\rightarrow$ 4 transitions towards the most spatially extended substructure within B1-E, the B1-E5 region, using the {\it Herschel Space Observatory}. A 3 and an 8 km s$^{-1}$ component were detected in both transitions, consistent with prior observations of lower rotational transitions. Using archival measurements of the $^{12}$CO 1 $\rightarrow$ 0, $^{12}$CO 3 $\rightarrow$ 2, $^{13}$CO 1 $\rightarrow$ 0, and $^{13}$CO 2 $\rightarrow$ 1 lines, SLEDs were created for both components and compared to the SLEDs predicted from three PDR codes: the \citet{Kaufman99}, \textsc{meudon}, and \textsc{kosma}-$\tau$ PDR codes. 

The ratios of the lower rotational transitions of the 3 km s$^{-1}$ component are well fit by PDR models with densities between 10$^3$ and 10$^4$ cm$^{-3}$, but these models underpredict the relative integrated intensity of the $^{12}$CO J = 6 $\rightarrow$ 5 line. The 8 km s$^{-1}$ component requires a slightly higher density, but the best-fitting PDR models can also reproduce the ratios of the $^{12}$CO J = 5 $\rightarrow$ 4 and lower transitions, while underpredicting the relative strength of the 6 $\rightarrow$ 5 line. The PDR models can also reasonably reproduce the absolute integrated intensities of the lower lines, but fail to simultaneously reproduce the 6 $\rightarrow$ 5 integrated intensity. The PDR models can also reproduce the archival low-J $^{13}$CO integrated intensities of both components.

We interpret this excess $^{12}$CO J = 6 $\rightarrow$ 5 emission as coming from a warm gas component within B1-E5 that comprises a small volume fraction of the region. We suggest that this emission is coming from gas heated by low-velocity shocks generated by the dissipation of turbulence within B1-E5, as predicted by \citet{Pon12Kaufman}. We show that the observed emission is consistent with the shock models of \citet{Pon12Kaufman} and find that the shocked gas has a volume filling factor of 0.15 per cent. We further calculate that the turbulent energy dissipation rate of the B1-E region is $3.5 \times 10^{32}$ erg s$^{-1}$ and that turbulent energy dissipation time-scale is only a factor of 3 smaller than the flow crossing time-scale for the B1-E region. 

\section*{ACKNOWLEDGEMENTS}
We would like to thank our anonymous referee for many useful changes to this paper. DJ acknowledges support from a Natural Sciences and Engineering Research Council (NSERC) Discovery Grant. This research has made use of the Smithsonian Astrophysical Observatory (SAO) / National Aeronautics and Space Administration's (NASA's) Astrophysics Data System (ADS). PC acknowledges the financial support of the European Research Council (ERC; project PALs 320620) and of successive rolling grants awarded by the UK Science and Technology Funding Council. HIFI has been designed and built by a consortium of institutes and university departments from across Europe, Canada, and the United States under the leadership of SRON Netherlands Institute for Space Research, Groningen, The Netherlands and with major contributions from Germany, France, and the United States. Consortium members are as follows. Canada: CSA, U.Waterloo; France: CESR, LAB, LERMA, IRAM; Germany: KOSMA, MPIfR, MPS; Ireland, NUI Maynooth; Italy: ASI, IFSI-INAF, Osservatorio Astrofisico di Arcetri-INAF; Netherlands: SRON, TUD; Poland: CAMK, CBK; Spain: Observatorio Astronómico Nacional (IGN), Centro de Astrobiología (CSIC-INTA); Sweden: Chalmers University of Technology -- MC2, RSS \& GARD; Onsala Space Observatory; Swedish National Space Board, Stockholm University -- Stockholm Observatory; Switzerland: ETH Zurich, FHNW; USA: Caltech, JPL, NHSC. This research has made use of the astro-ph archive. Some spectral line data were taken from the Spectral Line Atlas of Interstellar Molecules (SLAIM, available at http://www.splatalogue.net; F. J. Lovas, private communication; \citealt{Remijan07}). {\it Herschel} is an European Space Agency space observatory with science instruments provided by European-led Principal Investigator consortia and with important participation from NASA.

\bibliographystyle{mn2e}
\bibliography{ponbib}

\begin{thebibliography}{}
 \providecommand{\href}[2]{#2}

\bibitem[\protect\citeauthoryear{{Arce}, {Shepherd}, {Gueth}, {Lee},
  {Bachiller}, {Rosen} \& {Beuther}}{{Arce} et~al.}{2007}]{Arce07}
{Arce} H.~G.,  {Shepherd} D.,  {Gueth} F.,  {Lee} C.,  {Bachiller} R.,  {Rosen}
  A.,    {Beuther} H.,  2007, in {Reipurth} B.,  {Jewitt} D.,   {Keil} K.,
  eds, Protostars and Planets V. {Tucson, AZ: Univ. Arizona Press}, pp 245--260

\bibitem[\protect\citeauthoryear{{Arce}, {Borkin}, {Goodman}, {Pineda} \&
  {Beaumont}}{{Arce} et~al.}{2011}]{Arce11}
{Arce} H.~G.,  {Borkin} M.~A.,  {Goodman} A.~A.,  {Pineda} J.~E.,    {Beaumont}
  C.~N.,  2011, \apj, 742, 105

\bibitem[\protect\citeauthoryear{{Bachiller} \& {Cernicharo}}{{Bachiller} \&
  {Cernicharo}}{1986}]{Bachiller86}
{Bachiller} R.,  {Cernicharo} J.,  1986, \aap, 166, 283

\bibitem[\protect\citeauthoryear{{Bally}, {Walawender}, {Johnstone}, {Kirk} \&
  {Goodman}}{{Bally} et~al.}{2008}]{Bally08Walawender}
{Bally} J.,  {Walawender} J.,  {Johnstone} D.,  {Kirk} H.,    {Goodman} A.,
  2008, in {Reipurth} B.,  ed., Handbook of Star Forming Regions, Volume I.
  {San Francisco, CA: ASP Monograph Publications}, p.~308

\bibitem[\protect\citeauthoryear{{Basu} \& {Murali}}{{Basu} \&
  {Murali}}{2001}]{Basu01}
{Basu} S.,  {Murali} C.,  2001, \apj, 551, 743

\bibitem[\protect\citeauthoryear{{Caselli}, {Walmsley}, {Tafalla}, {Dore} \&
  {Myers}}{{Caselli} et~al.}{1999}]{Caselli99}
{Caselli} P.,  {Walmsley} C.~M.,  {Tafalla} M.,  {Dore} L.,    {Myers} P.~C.,
  1999, \apjl, 523, L165

\bibitem[\protect\citeauthoryear{{Enoch} et~al.,}{{Enoch}
  et~al.}{2006}]{Enoch06}
{Enoch} M.~L.  et~al., 2006, \apj, 638, 293

\bibitem[\protect\citeauthoryear{{Evans} II et~al.,}{{Evans}
  et~al.}{2009}]{Evans09}
{Evans} II N.~J.  et~al., 2009, \apjs, 181, 321

\bibitem[\protect\citeauthoryear{{Falgarone}, {Pety} \&
  {Hily-Blant}}{\protect\mniiiauthor{Falgarone09}{{Falgarone}, {Pety} \&
  {Hily-Blant}}{{Falgarone} et~al.}}{2009}]{Falgarone09}
{Falgarone} E.,  {Pety} J.,    {Hily-Blant} P.,  2009, \aap, 507, 355

\bibitem[\protect\citeauthoryear{{Falgarone} et~al.,}{{Falgarone}
  et~al.}{2010}]{Falgarone10Ossenkopf}
{Falgarone} E.  et~al., 2010, \aap, 518, L118

\bibitem[\protect\citeauthoryear{{Federman}, {Glassgold} \&
  {Kwan}}{\protect\mniiiauthor{Federman79}{{Federman}, {Glassgold} \&
  {Kwan}}{{Federman} et~al.}}{1979}]{Federman79}
{Federman} S.~R.,  {Glassgold} A.~E.,    {Kwan} J.,  1979, \apj, 227, 466

\bibitem[\protect\citeauthoryear{{Fontani}, {Giannetti}, {Beltr{\'a}n},
  {Dodson}, {Rioja}, {Brand}, {Caselli} \& {Cesaroni}}{{Fontani}
  et~al.}{2012}]{Fontani12}
{Fontani} F.,  {Giannetti} A.,  {Beltr{\'a}n} M.~T.,  {Dodson} R.,  {Rioja} M.,
   {Brand} J.,  {Caselli} P.,    {Cesaroni} R.,  2012, \mnras, 423, 2342

\bibitem[\protect\citeauthoryear{{Gammie} \& {Ostriker}}{{Gammie} \&
  {Ostriker}}{1996}]{Gammie96}
{Gammie} C.~F.,  {Ostriker} E.~C.,  1996, \apj, 466, 814

\bibitem[\protect\citeauthoryear{{Glover} \& {Clark}}{{Glover} \&
  {Clark}}{2012}]{Glover12a}
{Glover} S.~C.~O.,  {Clark} P.~C.,  2012, \mnras, 421, 9

\bibitem[\protect\citeauthoryear{{Glover} \& {Mac Low}}{{Glover} \& {Mac
  Low}}{2011}]{Glover11}
{Glover} S.~C.~O.,  {Mac Low} M.-M.,  2011, \mnras, 412, 337

\bibitem[\protect\citeauthoryear{{Godard}, {Falgarone} \& {Pineau Des
  For{\^e}ts}}{\protect\mniiiauthor{Godard09}{{Godard}, {Falgarone} \& {Pineau
  Des For{\^e}ts}}{{Godard} et~al.}}{2009}]{Godard09}
{Godard} B.,  {Falgarone} E.,    {Pineau Des For{\^e}ts} G.,  2009, \aap, 495,
  847

\bibitem[\protect\citeauthoryear{{Goodman}, {Pineda} \&
  {Schnee}}{\protect\mniiiauthor{Goodman09Pineda}{{Goodman}, {Pineda} \&
  {Schnee}}{{Goodman} et~al.}}{2009}]{Goodman09Pineda}
{Goodman} A.~A.,  {Pineda} J.~E.,    {Schnee} S.~L.,  2009, \apj, 692, 91

\bibitem[\protect\citeauthoryear{{Hacar}, {Tafalla}, {Kauffmann} \&
  {Kov{\'a}cs}}{{Hacar} et~al.}{2013}]{Hacar13}
{Hacar} A.,  {Tafalla} M.,  {Kauffmann} J.,    {Kov{\'a}cs} A.,  2013, \aap,
  554, A55

\bibitem[\protect\citeauthoryear{{Hennebelle} \& {Andr{\'e}}}{{Hennebelle} \&
  {Andr{\'e}}}{2013}]{Hennebelle13Andre}
{Hennebelle} P.,  {Andr{\'e}} P.,  2013, \aap, 560, A68

\bibitem[\protect\citeauthoryear{{Henshaw}, {Caselli}, {Fontani},
  {Jim{\'e}nez-Serra}, {Tan} \& {Hernandez}}{{Henshaw}
  et~al.}{2013}]{Henshaw13}
{Henshaw} J.~D.,  {Caselli} P.,  {Fontani} F.,  {Jim{\'e}nez-Serra} I.,  {Tan}
  J.~C.,    {Hernandez} A.~K.,  2013, \mnras, 428, 3425

\bibitem[\protect\citeauthoryear{{Hernandez}, {Tan}, {Caselli}, {Butler},
  {Jim{\'e}nez-Serra}, {Fontani} \& {Barnes}}{{Hernandez}
  et~al.}{2011}]{Hernandez11Caselli}
{Hernandez} A.~K.,  {Tan} J.~C.,  {Caselli} P.,  {Butler} M.~J.,
  {Jim{\'e}nez-Serra} I.,  {Fontani} F.,    {Barnes} P.,  2011, \apj, 738, 11

\bibitem[\protect\citeauthoryear{{Hollenbach}, {Kaufman}, {Neufeld}, {Wolfire}
  \& {Goicoechea}}{{Hollenbach} et~al.}{2012}]{Hollenbach12}
{Hollenbach} D.,  {Kaufman} M.~J.,  {Neufeld} D.,  {Wolfire} M.,
  {Goicoechea} J.~R.,  2012, \apj, 754, 105

\bibitem[\protect\citeauthoryear{{J{\o}rgensen}, {Johnstone}, {Kirk} \&
  {Myers}}{{J{\o}rgensen} et~al.}{2007}]{Jorgensen07}
{J{\o}rgensen} J.~K.,  {Johnstone} D.,  {Kirk} H.,    {Myers} P.~C.,  2007,
  \apj, 656, 293

\bibitem[\protect\citeauthoryear{{Kaufman} \& {Neufeld}}{{Kaufman} \&
  {Neufeld}}{1996a}]{Kaufman96I}
{Kaufman} M.~J.,  {Neufeld} D.~A.,  1996a, \apj, 456, 250

\bibitem[\protect\citeauthoryear{{Kaufman} \& {Neufeld}}{{Kaufman} \&
  {Neufeld}}{1996b}]{Kaufman96II}
{Kaufman} M.~J.,  {Neufeld} D.~A.,  1996b, \apj, 456, 611

\bibitem[\protect\citeauthoryear{{Kaufman}, {Wolfire}, {Hollenbach} \&
  {Luhman}}{{Kaufman} et~al.}{1999}]{Kaufman99}
{Kaufman} M.~J.,  {Wolfire} M.~G.,  {Hollenbach} D.~J.,    {Luhman} M.~L.,
  1999, \apj, 527, 795

\bibitem[\protect\citeauthoryear{{Kazandjian}, {Meijerink}, {Pelupessy},
  {Israel} \& {Spaans}}{{Kazandjian} et~al.}{2012}]{Kazandjian12}
{Kazandjian} M.~V.,  {Meijerink} R.,  {Pelupessy} I.,  {Israel} F.~P.,
  {Spaans} M.,  2012, \aap, 542, A65

\bibitem[\protect\citeauthoryear{{Kazandjian}, {Meijerink}, {Pelupessy},
  {Israel} \& {Spaans}}{{Kazandjian} et~al.}{2014}]{Kazandjian14}
{Kazandjian} M.~V.,  {Meijerink} R.,  {Pelupessy} I.,  {Israel} F.~P.,
  {Spaans} M.,  2014, ArXiv e-prints:1403.7000

\bibitem[\protect\citeauthoryear{{Kirk}, {Johnstone} \& {Di
  Francesco}}{\protect\mniiiauthor{Kirk06}{{Kirk}, {Johnstone} \& {Di
  Francesco}}{{Kirk} et~al.}}{2006}]{Kirk06}
{Kirk} H.,  {Johnstone} D.,    {Di Francesco} J.,  2006, \apj, 646, 1009

\bibitem[\protect\citeauthoryear{{Larson}}{{Larson}}{1981}]{Larson81}
{Larson} R.~B.,  1981, \mnras, 194, 809

\bibitem[\protect\citeauthoryear{{Le Petit}, {Nehm{\'e}}, {Le Bourlot} \&
  {Roueff}}{{Le Petit} et~al.}{2006}]{LePetit06}
{Le Petit} F.,  {Nehm{\'e}} C.,  {Le Bourlot} J.,    {Roueff} E.,  2006, \apjs,
  164, 506

\bibitem[\protect\citeauthoryear{{Lesaffre}, {Pineau des For{\^e}ts}, {Godard},
  {Guillard}, {Boulanger} \& {Falgarone}}{{Lesaffre} et~al.}{2013}]{Lesaffre13}
{Lesaffre} P.,  {Pineau des For{\^e}ts} G.,  {Godard} B.,  {Guillard} P.,
  {Boulanger} F.,    {Falgarone} E.,  2013, \aap, 550, A106

\bibitem[\protect\citeauthoryear{{Lord}, {Hollenbach}, {Haas}, {Rubin},
  {Colgan} \& {Erickson}}{{Lord} et~al.}{1996}]{Lord96}
{Lord} S.~D.,  {Hollenbach} D.~J.,  {Haas} M.~R.,  {Rubin} R.~H.,  {Colgan}
  S.~W.~J.,    {Erickson} E.~F.,  1996, \apj, 465, 703

\bibitem[\protect\citeauthoryear{{Mac Low}}{{Mac Low}}{1999}]{MacLow99}
{Mac Low} M.,  1999, \apj, 524, 169

\bibitem[\protect\citeauthoryear{{Mac Low}, {Klessen}, {Burkert} \&
  {Smith}}{{Mac Low} et~al.}{1998}]{MacLow98}
{Mac Low} M.,  {Klessen} R.~S.,  {Burkert} A.,    {Smith} M.~D.,  1998,
  Physical Review Letters, 80, 2754

\bibitem[\protect\citeauthoryear{{Mathis}, {Mezger} \&
  {Panagia}}{\protect\mniiiauthor{Mathis83}{{Mathis}, {Mezger} \&
  {Panagia}}{{Mathis} et~al.}}{1983}]{Mathis83}
{Mathis} J.~S.,  {Mezger} P.~G.,    {Panagia} N.,  1983, \aap, 128, 212

\bibitem[\protect\citeauthoryear{{Ostriker}, {Stone} \&
  {Gammie}}{\protect\mniiiauthor{Ostriker01}{{Ostriker}, {Stone} \&
  {Gammie}}{{Ostriker} et~al.}}{2001}]{Ostriker01}
{Ostriker} E.~C.,  {Stone} J.~M.,    {Gammie} C.~F.,  2001, \apj, 546, 980

\bibitem[\protect\citeauthoryear{{Ott}}{{Ott}}{2010}]{Ott10}
{Ott} S.,  2010, in {Mizumoto} Y.,  {Morita} K.-I.,   {Ohishi} M.,  eds,
  Astronomical Society of the Pacific Conference Series Vol. 434, Astronomical
  Data Analysis Software and Systems XIX. p.~139

\bibitem[\protect\citeauthoryear{{Padoan} \& {Nordlund}}{{Padoan} \&
  {Nordlund}}{1999}]{Padoan99}
{Padoan} P.,  {Nordlund} {\AA}.,  1999, \apj, 526, 279

\bibitem[\protect\citeauthoryear{{Pilbratt} et~al.,}{{Pilbratt}
  et~al.}{2010}]{Pilbratt10}
{Pilbratt} G.~L.  et~al., 2010, \aap, 518, L1

\bibitem[\protect\citeauthoryear{{Pineda}, {Caselli} \&
  {Goodman}}{\protect\mniiiauthor{Pineda08}{{Pineda}, {Caselli} \&
  {Goodman}}{{Pineda} et~al.}}{2008}]{Pineda08}
{Pineda} J.~E.,  {Caselli} P.,    {Goodman} A.~A.,  2008, \apj, 679, 481

\bibitem[\protect\citeauthoryear{{Pineda}, {Goodman}, {Arce}, {Caselli},
  {Foster}, {Myers} \& {Rosolowsky}}{{Pineda} et~al.}{2010}]{Pineda10}
{Pineda} J.~E.,  {Goodman} A.~A.,  {Arce} H.~G.,  {Caselli} P.,  {Foster}
  J.~B.,  {Myers} P.~C.,    {Rosolowsky} E.~W.,  2010, \apjl, 712, L116

\bibitem[\protect\citeauthoryear{{Pon}, {Johnstone} \&
  {Kaufman}}{\protect\mniiiauthor{Pon12Kaufman}{{Pon}, {Johnstone} \&
  {Kaufman}}{{Pon} et~al.}}{2012}]{Pon12Kaufman}
{Pon} A.,  {Johnstone} D.,    {Kaufman} M.~J.,  2012, \apj, 748, 25

\bibitem[\protect\citeauthoryear{{Remijan}, {Markwick-Kemper} \& {ALMA Working
  Group on Spectral Line
  Frequencies}}{\protect\mniiiauthor{Remijan07}{{Remijan}, {Markwick-Kemper} \&
  {ALMA Working Group on Spectral Line Frequencies}}{{Remijan}
  et~al.}}{2007}]{Remijan07}
{Remijan} A.~J.,  {Markwick-Kemper} A.,    {ALMA Working Group on Spectral Line
  Frequencies} 2007, in American Astronomical Society Meeting Abstracts. p.~963

\bibitem[\protect\citeauthoryear{{Ridge} et~al.,}{{Ridge}
  et~al.}{2006a}]{Ridge06DiFrancesco}
{Ridge} N.~A.  et~al., 2006a, \aj, 131, 2921

\bibitem[\protect\citeauthoryear{{Ridge}, {Schnee}, {Goodman} \&
  {Foster}}{{Ridge} et~al.}{2006b}]{Ridge06Schnee}
{Ridge} N.~A.,  {Schnee} S.~L.,  {Goodman} A.~A.,    {Foster} J.~B.,  2006b,
  \apj, 643, 932

\bibitem[\protect\citeauthoryear{{Roelfsema} et~al.,}{{Roelfsema}
  et~al.}{2012}]{Roelfsema12}
{Roelfsema} P.~R.  et~al., 2012, \aap, 537, A17

\bibitem[\protect\citeauthoryear{{R{\"o}llig} \& {Ossenkopf}}{{R{\"o}llig} \&
  {Ossenkopf}}{2013}]{Rollig13Ossenkopf}
{R{\"o}llig} M.,  {Ossenkopf} V.,  2013, \aap, 550, A56

\bibitem[\protect\citeauthoryear{{R{\"o}llig}, {Ossenkopf}, {Jeyakumar},
  {Stutzki} \& {Sternberg}}{{R{\"o}llig} et~al.}{2006}]{Rollig06}
{R{\"o}llig} M.,  {Ossenkopf} V.,  {Jeyakumar} S.,  {Stutzki} J.,
  {Sternberg} A.,  2006, \aap, 451, 917

\bibitem[\protect\citeauthoryear{{R{\"o}llig} et~al.,}{{R{\"o}llig}
  et~al.}{2007}]{Rollig07}
{R{\"o}llig} M.  et~al., 2007, \aap, 467, 187

\bibitem[\protect\citeauthoryear{{R{\"o}llig}, {Szczerba}, {Ossenkopf} \&
  {Gl{\"u}ck}}{{R{\"o}llig} et~al.}{2013}]{Rollig13Szczerba}
{R{\"o}llig} M.,  {Szczerba} R.,  {Ossenkopf} V.,    {Gl{\"u}ck} C.,  2013,
  \aap, 549, A85

\bibitem[\protect\citeauthoryear{{Sadavoy} et~al.,}{{Sadavoy}
  et~al.}{2012}]{Sadavoy12}
{Sadavoy} S.~I.  et~al., 2012, \aap, 540, A10

\bibitem[\protect\citeauthoryear{{Sadavoy} et~al.,}{{Sadavoy}
  et~al.}{2014}]{Sadavoy14}
{Sadavoy} S.~I.  et~al., 2014, \apjl, 787, L18

\bibitem[\protect\citeauthoryear{{Sch{\"o}ier}, {van der Tak}, {van Dishoeck}
  \& {Black}}{{Sch{\"o}ier} et~al.}{2005}]{Schoier05}
{Sch{\"o}ier} F.~L.,  {van der Tak} F.~F.~S.,  {van Dishoeck} E.~F.,    {Black}
  J.~H.,  2005, \aap, 432, 369

\bibitem[\protect\citeauthoryear{{Solomon}, {Rivolo}, {Barrett} \&
  {Yahil}}{{Solomon} et~al.}{1987}]{Solomon87}
{Solomon} P.~M.,  {Rivolo} A.~R.,  {Barrett} J.,    {Yahil} A.,  1987, \apj,
  319, 730

\bibitem[\protect\citeauthoryear{{Sternberg} \& {Dalgarno}}{{Sternberg} \&
  {Dalgarno}}{1989}]{Sternberg89}
{Sternberg} A.,  {Dalgarno} A.,  1989, \apj, 338, 197

\bibitem[\protect\citeauthoryear{{Stone}, {Ostriker} \&
  {Gammie}}{\protect\mniiiauthor{Stone98}{{Stone}, {Ostriker} \&
  {Gammie}}{{Stone} et~al.}}{1998}]{Stone98}
{Stone} J.~M.,  {Ostriker} E.~C.,    {Gammie} C.~F.,  1998, \apjl, 508, L99

\bibitem[\protect\citeauthoryear{{Sun}, {Kramer}, {Ossenkopf}, {Bensch},
  {Stutzki} \& {Miller}}{{Sun} et~al.}{2006}]{Sun06}
{Sun} K.,  {Kramer} C.,  {Ossenkopf} V.,  {Bensch} F.,  {Stutzki} J.,
  {Miller} M.,  2006, \aap, 451, 539

\bibitem[\protect\citeauthoryear{{Tielens}}{{Tielens}}{2005}]{Tielens05}
{Tielens} A.~G.~G.~M.,  2005, {The Physics and Chemistry of the Interstellar
  Medium}.
Cambridge, UK: Cambridge University Press

\bibitem[\protect\citeauthoryear{{Wolfire}, {Tielens} \&
  {Hollenbach}}{\protect\mniiiauthor{Wolfire90}{{Wolfire}, {Tielens} \&
  {Hollenbach}}{{Wolfire} et~al.}}{1990}]{Wolfire90}
{Wolfire} M.~G.,  {Tielens} A.~G.~G.~M.,    {Hollenbach} D.,  1990, \apj, 358,
  116

\bibitem[\protect\citeauthoryear{{de Graauw} et~al.,}{{de Graauw}
  et~al.}{2010}]{deGraauw10}
{de Graauw} T.  et~al., 2010, \aap, 518, L6

\bibitem[\protect\citeauthoryear{{van der Tak}, {Black}, {Sch{\"o}ier},
  {Jansen} \& {van Dishoeck}}{{van der Tak} et~al.}{2007}]{Vandertak07}
{van der Tak} F.~F.~S.,  {Black} J.~H.,  {Sch{\"o}ier} F.~L.,  {Jansen} D.~J.,
    {van Dishoeck} E.~F.,  2007, \aap, 468, 627

\end{thebibliography}

\end{document}